\documentclass[aps,prd,superscriptaddress,showpacs,twocolumn,floatfix]{revtex4}
\usepackage{graphicx}

\newcommand{\ba}{\begin{eqnarray}}
\newcommand{\ea}{\end{eqnarray}}
\newcommand{\beqs}{\begin{eqnarray}}
\newcommand{\eeqs}{\end{eqnarray}}

\begin{document}
\title{Models of  parton distributions and the description of
       form factors of nucleon.}

\author{  O.V. Selyugin\footnote{selugin@theor.jinr.ru} } 
\address{\it Bogoliubov Laboratory of Theoretical Physics, \\
Joint Institute for Nuclear Research,
141980 Dubna, Moscow region, Russia }

\pacs{{12.38.Lg} 
      {13.40.Gp}, 
      {13.60.Fz}, 
      {14.20.Dh}, 
     } 
\begin{abstract}
The comparative analysis of different sets of the  parton distribution functions (PDFs),
    based on the description of the whole sets of experimental data
    of electromagnetic form factors of the proton and neutron, is made
    in the framework of the model  of t dependence of the generalized parton distributions (GPDs)
     with minimum free parameters and  some extending variants of the  model.
    In some cases, a large difference in the description of electromagnetic form factors of  nucleons
    with using the different sets of  PDF  are found out.
    The different variants of the flavor dependence of the up and down quark form factors are presented
    and discussed.
    The gravitation form factors, obtained with the different sets of  PDF, are also
    calculated  and the anomalous gravimagnetic moment is compared with the equivalence principle.
    The calculations of the differential cross sections of the real Compton scattering are presented.
\end{abstract}

\maketitle %

\section{\label{sec:intro}Introduction}

   The structure of  nucleons is the most intriguing problem of the old and new physics.
   In the first place, it is connected with the electromagnetic structure of the nucleon
   which can be obtained from the electron-hadron elastic scattering.
   In the Born approximation,
       the Feynman amplitude for the elastic electron-proton scattering \cite{Lomon} is
  \ba
M_{ep\rightarrow ep} = \frac{1}{q^2}[e \bar{u}(k_2)\gamma^{\mu} u(k_{1}][e\bar{U}(p_{2}\Gamma_{\mu}(p_{1},p_{2})U(p_{1}],
 \ea
   where $u$ and $U$ are the electron and nucleon a Dirac spinors,
\ba
\Gamma^{\mu} = F_{1}(t)\gamma^{\mu} \ + \  F_{2}(t) \frac{i\sigma^{\mu\nu} q_{\nu}}{2 m},
\ea
   where $m$ is the nucleon mass, $\kappa$ is the anomalous part
   of the magnetic moment and $t=-q^2 =-(p-p^{\prime})^2$ is the square of the momentum transfer of the nucleon.

   The functions $F_{1}(t)$ and $F_{2}(t)$ are named  the Dirac and Pauli form factors, which depend upon the nucleon structure.
   The normalization of the form factors  \cite{QA12}  is given by
   \ba
   F_{1}^{p}(t=0)=1, \ \ \ F_{2}^{p}(t=0)=\kappa_{p}=1.793
   \ea
   for the proton and
   \ba
   F_{1}^{n}(t=0)=0, \ \ \ F_{2}^{n}(t=0)=\kappa_{n}=-1.913
   \ea
   for the neutron.

 Two important combinations of the  Dirac and Pauli form factors
are the so-called Sachs form factors \cite{Ernst60,Sachs62}.
   In the Breit frame the current is separated into the
   electric and magnetic contributions \cite{Kelly-02}
   \ba
  \bar{u}(p^{\prime},s^{\prime}) \Gamma^{\mu} u(p,s) = \chi^{\dag}_{s^{\prime}}
  \left( G_{E}(t) + \frac{i \vec{\sigma} \times \vec{q_{B}}}{2 m} G_{M}(t)  \right) \chi_{s},
  \ea
 where $\chi_{s}$ is the two-component of the Pauli spinor,
 $ G_{E}(t)$ and $G_{M}(t)$ are the Sachs form factors given by
 \ba
  G_{E}^{p/n}(t) = F_{1}^{p/n}(t)-\tau  F_{2}^{p/n}(t),
  \ea
  \ba
  G_{M}^{p/n}(t) = F_{1}^{p/n}(t) +  F_{2}^{p/n}(t),
  \ea
  where $\tau=t/(4 M^{2})$.
Their three-dimensional Fourier transform  provides the electric charge density
 and the magnetic current density distribution \cite{Sachs62}.
  Those form factors can be extracted from  experimental data on the elastic electron-nucleon
  scattering by the Rosenbluth method or from the polarization electron proton elastic scattering.

Some experiments were based on the Rosenbluth formula \cite{Rosenbluth}
  \ba
\frac{d \sigma}{d \Omega}=\frac{\sigma_{Mott}}{\epsilon (1+\tau)}
 [\tau G^2_{M}(t) + \epsilon G^2_{E}(t)],
 \ea
 where  $\tau= Q^2/4M^2_p$ and $\epsilon = [1+2(1+\tau) \tan^2(\theta_e/2)]^{-1}$
 is the measure of the virtual photon polarization.
 Early experiments at modest $t$, based on the Rosenbluth separation method, suggested
 that the scaling behavior of both the proton form factors and the neutron magnetic form factor
  approximately described by a dipole form
\ba
G^{p}_{E}(t) \approx \frac{G^{p}_{M}(t)}{\mu_{p}} \approx \frac{G^{n}_{M}(t)}{\mu_{n}}
\approx G_{D}(t) = \frac{\Lambda^4}{(\Lambda^{2} -t)^2},
\ea
which leads to
\ba
 F_{1}^{D} (t) = \frac{4M_{p}^{2} - t \mu_{p}}{4M_{p}^{2} - t }
G_{D}(t);
\ea
\ba
F_{2}^{D} (t) =  \frac{4 k_{p} M_{p}^{2}}{ (4 M^{2}_{p} - t]} G_{D}(t);
\ea
with $\Lambda^2= 0.71$ GeV$^{2}$.

   Recently,  better data have been obtained 
 by using  the polarization method \cite{Akhiezer,Arnold}.
 Measuring both transverse
 and longitudinal components of the recoil proton polarization in the electron
 scattering plane,  the data on the ratio
 \ba
 \frac{G^{p}_{E}(t) }{ G^{p}_{M}(t) } = -\frac{ P_t }{ P_l } \frac{ E+E^{'}}{2 M_{p} } \tan(\theta/2)
 \ea
 were obtained. These data manifested a strong deviation from the
 scaling law and, consequently, disagreement with data obtained by the Rosenbluth technique.
 The results consist in   an almost linear decrease of $G^{p}_{E} / G^{p}_{M}$.
 There were attempts to solve the problem by inclusion of additional
 radiative correction terms related to two-photon exchange approximations ( for example,
 \cite{Guichon}).
 In recent works \cite{Kuraev1,Kuraev2}, the box amplitude is calculated
 when the intermediate state is a proton or the $\Delta$ resonance.
 The results of the numerical estimation show that the present calculation of
 radiative corrections can bring into  better agreement
  the conflicting experimental results
 on proton electromagnetic form factors.
 Note, however, that the  data of a Rosenbluth measurement of the proton form factors
 at $Q^2 = 4. \ $ GeV$^2$ \cite{Qat05} lie so high
  that they require very large corrections to move them down to meet the polarization data.

   In the parton language, the hadron structure
    can be described by the parton distribution functions (PDFs).
   In the quantum chromodynamics (QCD)
   it can be presented by gluons and quarks. Practically, all modern descriptions of the high-energy experiments are based on  some PDFs of the hadrons.
   To our regret, at the present time  PDFs cannot be calculated from the first principles. They are determined by the modeling of the dip inelastic processes, including modern
   physical results obtained at the LHC. Including the new experimental results leads to the change of the parameters
   of the PDF model description. The different forms of  PDF were proposed during
    the last 15 years.
   Now all these models give a sufficiently  good description of the high-energy experimental data
   on the dip inelastic processes.


   The hadronic current as a sum of quark currents can be decomposed
   into the Pauli and Dirac form factors of the nucleon with the
   flavor quark components \cite{Gates}
   \ba
   F_{1,2}^{u}(t)&=& 2 F_{1,2}^{p}(t) + F_{1,2}^{n}(t); \nonumber \\
   F_{1,2}^{d}(t)&=& F_{1,2}^{p}(t) + 2 F_{1,2}^{n}(t),
   \ea
 with the normalization $F_{1}^{u}(t=0)= 2$, $F_{2}^{u}(t=0)= \kappa_{u}$, and
 $F_{1}^{d}(t=0)= 1$, $F_{2}^{d}(t=0)= \kappa_{d}$, where the anomalous magnetic moments
 for the $u$ and $d$ quarks are $\kappa_{u}=2 \kappa_{p} + \kappa_{n}=1.673$
 and $\kappa_{d}=\kappa_{p} + 2 \kappa_{n}=-2.033$.

The next step in the development of the picture of the hadron
 was made by introducing the nonforward structure functions, general parton distributions (GPDs)
   \cite{Muller94,Ji97,R97}  with the spin-independent  $H_{q}(x,\xi,t) $
   and the spin-dependent $E_{q}(x,\xi,t)$ parts.
  Generally,  GPDs depend on the momentum transfer $t$, and the average momentum fraction
  $x=0.5 (x_i + x_f)$ of the active quark, and the skewness parameter
  $2 \xi=x_f - x_i$  measures the longitudinal momentum transfer.
  One can choose the special case $\xi=0$ of the nonforward parton densities \cite{R98} ${ \cal{F}}^{a}_{ \xi } (x ; t)$
   for which the emitted and reabsorbed partons carry the same momentum fractions:
\ba
{\cal{H}}^{q} (x,t) \ = \ H^{q}(x,\xi=0,t)  \ - \ H^{\bar q} (-x,\xi=0,t),
\ea
\ba
{\cal{E}}^{q} (x,t) \ = \  E^{q}(x,\xi=0,t) \ - \ E^{\bar q} (-x,\xi=0,t).
\ea

 Some of the advantages of  GPDs were presented by the sum rules \cite{Ji97}
   which impose the connections of GPDs with the standard electromagnetic hadron form factors
  \ba
 F_{1}^q (t) = \int^{1}_{0} \ dx  \ {\cal{ H}}^{q} (x,\xi=0, t),
\ea
\ba
 F_{2}^q (t) = \int^{1}_{0} \ dx \  {\cal{E}}^{q} (x, \xi=0, t).
\ea
Non-forward parton densities also provide information
about the distribution of the parton in the impact parameter space \cite{Burk00}, which
is connected with the $t$ dependence of GPDs.
Now we cannot obtain this dependence from the first principles,
but it must be obtained from the phenomenological description with GPDs
of the nucleon electromagnetic form factors.

The obtaining of the true $t$ dependence of GPDs in a straightforward way
  from the analysis of the dip inelastic processes meets  many problems.
  Such analysis requires to take into account
  the gluon and sea contributions and many assumptions about these processes
  (see, for example, \cite{Kumer10,Kumer11}).
    The additional $\xi$ dependence and, in most part, bound on the size of $x$ and $t$  create  a wide corridor for the $t$ dependence of GPDs \cite{Guidal13}.   

   Note, that in some  works the factorization form of  GPDs  was used.
   The factorization supposes that all $x$ dependence of GPDs is concentrated in
    PDFs and all $t$ dependence is concentrated in the Regge-like exponential form.
    Such a factorization form cannot describe the corresponding electromagnetic form factors  in a wide region of the momentum transfer,
    as we know that they can have the approximately exponential form only at small momentum transfer.

Many different forms
  of the $t$ dependence of GPDs were proposed.
      There are two approaches to the GPDs:
    1) the factorization form, where the t dependence is taken in the
    simple factorized {\it Ansatz}  with Regge-like form for the t dependence of GPDs
    \cite{Vander01,Boffi07}, and
    (2) the nonfactorization form,  where the function with the $t$ dependence has
    some complicated form  of $x$  \cite{R04,Guidal13}  
  \ba
{\cal{H}}^{q} (x,t) \  \sim  q(x) e^{  f(x)^{q} \ t }.
\ea
       In  \cite{R04},  $f(x,t)$  was taken in two forms
  \ba
       a) (R1) f(x,t)&=&-t \ Ln(x);  \\
       b) (R2) f(x,t)&=&-t (1-x) \ Ln(x) .
       \ea
     In the last case they made a qualitative analysis of the nucleon form factors.

   In the quark diquark model \cite{Liuti1,Liuti2}, the form of  GPDs
   consists of three parts - PDFs, function distribution and the Regge-like function,
\ba
H(E)_{q}(x,t)=N_{q} \ G_{M_{x}^{I.II}}^{\lambda^{I.II}}(x,t) \ R_{Pq}^{\alpha_{q} \alpha_{q}^{\prime}}(x,t).
 \ea
  The parameters have the flavor dependence for  all three parts.
 In other works (see, e.g., \cite{Kroll04,Kroll13})
  the description of the $t$ dependence of  GPDs  was developed
  in a complicated picture using the exponential with polynomial forms with respect to $x$
  with
  \ba
  f^{q}(x) = A_{q} \ (1-x)^n Log(1/x) + B_{q}(1-x)^{n-1} +C_{q},
\ea
 where $n=3$ or $n=2$ in the different variants and the coefficients $A_{q}$, $B_{q}$,
 and $C_{q}$ are the flavor dependence.

    Note that  in \cite{Yuan03},  
    it was shown that at  large $x  \rightarrow 1$
    and momentum transfer the behavior of GPDs
  requires a larger power of $(1-x)^{n}$ in the  $t$-dependent exponent
\ba
{\cal{H}}^{q} (x,t) \  \sim  exp[ a \ (1-x)^n \ t ] \ q(x); \  \  \
   n \geq 2 .
   \label{n2}
\ea
It was noted that $n=2$ naturally leads to the Drell-Yan-West duality
 between parton distributions at large $x$ and the form factors.

       The existing experimental data of DVCS/DVMP  of HERMES and JLab  are
       obtained  only on some bins of $x_i,t_i$ at small $x$ and $t$.
       That and many different {\it Ansatze}
       and assumptions in the models of GPDs including the necessity to take into account
      the twist two and three contributions to the DVCS amplitude \cite{Anikin11}
       do not allow one to determine the corresponding $t$ dependence of GPDs.
       The model independent analysis of these data leads to the large uncertainty in the
       definition of GPDs parts \cite{Guidal09,Guidal10}.
       So, in our work we used an {\it Ansatz} with minimum free parameters
        based of some theoretical results
       and compared its form  with the complete sets of the experimental data on
       the electromagnetic form factors of the nucleons in  the region of small and large of $t$
       and using the different PDF sets in a wide region of $x$.
       Then we intend to use the obtained form of the electromagnetic and gravimagnetic
       form factor to describe the elastic hadron scattering in a wide region of the energy
       and momentum transfer.

   The hadron structure in the form of the form factors is used in the
   different models of the elastic hadron scattering \cite{Rev-LHC}.
   The new data of the TOTEM Collaboration \cite{TOTEM-1395,TOTEM-11} show that none of the model predictions can
   describe the high-energy elastic cross sections.
  The one of the main problems of the dynamical models is the form factors of the hadrons. In most part, the model is based
  on the assumption that the strong form factors correlate with the electromagnetic form factors. In practice, the models
  use some phenomenological forms of the form factors with the parameters determined by the fit of the experimental data of the hadron elastic scattering.
  In some works \cite{Miettinen,Valin},  the idea was introduced that the strong form factors can be proportional to the
  matter distribution of the hadrons. In \cite{M1}, the model was developed with the two forms of the form factors
  - one is the exact  electromagnetic form factors and the second is proportional to the matter distribution of the hadron.
  Both form factors were obtained from the General Parton distributions (GPDs), which are based on the parton distributions (PDF)
  obtained from the data on the dip inelastic scattering.  The model used the old PDF obtained in \cite{MRST02}.
  In the framework of the model, the good description of the high-energy  of the proton-proton and proton-antiproton
  elastic scattering was obtained only with 3 high-energy fitting parameters.
   The question arises how the different
  PDF sets describe the electromagnetic form factor of the hadrons.
 For that, we made for the first time the numerical simultaneous fits of all available experimental data on the proton and neutron electromagnetic form factors.
       In the framework of our model of the $t$ dependence of GPDs we made for the first time
        the comparative analysis of 24 sets of the PDFs of the
       different Collaborations and compared the obtained fitting parameters of $t$ dependence
       for the different PDFs. This allows us to determine the true size of our fitting parameters
        independently of the form of PDFs to determine the form of the electromagnetic $F_{1}(t)$
        and gravimagnetic $A_{gr}(t)$ form factors of the nucleons.

  In Secs. $II$ and $III$, we look through the different forms of the GPD and PDF sets
  of the different Collaborations.  In  Secs. $IV$, the fitting of a wide set of experimental data on the electromagnetic form factors of the proton and neutron with the different sets of PDF
  are carried out. In  Secs. $V$, the analysis of the flavor dependence of the separate parts of the
  electromagnetic form factors is given. The second moments of GPDs and
   the  corresponding
   gravimagnetic form factors are obtained and discussed in  Secs. VI.
   In  Secs. VII we present our calculations of the differential cross section of the real Compton scattering.

\section{The descriptions of the electromagnetic form factors}

The electromagnetic form factors can be represented as first  moments of GPDs
following from the sum rules \cite{Ji97}. 
 We introduced a simple form for this
 $t $ dependence~\cite{ST-PRDGPD} based on the original Gaussian form corresponding to that
 of the wave function of the hadron. It satisfies the conditions of nonfactorization,
 introduced in \cite{R98,R04}, and the  condition, Eq.(\ref{n2}), on the power of $(1-x)^n $
 in the exponential form of the $t $ dependence.

Let us modify the original Gaussian {\it Ansatz} in order to incorporate
the observations of \cite{R98} and \cite{Burk04} and choose
 the $t$ dependence of  GPDs in the usual form \cite{ST-PRDGPD}
\ba
{\cal{H}}^{q} (x,t)=  \ g^{q}(x) \ e^{f^{q}(x) \ t},
\label{Hbv}
\ea
\ba
{\cal{E}}^{q} (x,t)= \ g^{q}(x)\ g_{e}^{q}(x) \ e^{f^{q}(x) \ t},
\label{Ebv}
\ea
 with  
 \ba
 f^{q}(x)= && 2 \ \alpha_{H,E} \frac{(1-x)^{p_1}}{(x_0+x)^{p_2}}, 	
\label{fbv}
 \ea
 with $p_1=2$, $p_2 = 0.4 \div 0.5$ and $x_0 \approx 0$. In this case, the functions $f^{q}(x)$ are
  independent of the flavor of quarks.
  The additional function $g_{e}^{q}(x) $ was taken from the  corresponding work \cite{R04} in the form
  $(1-x)^{e_{q}}$ with $e_{u}=1.52$ for the $u$ quark and $e_{d}=0.31$ for the $d$ quark.
 With this form and  PDFs obtained in  \cite{MRST02}, we get the qualitatively good descriptions of the electromagnetic form factors of the proton and neutron \cite{ST-PRDGPD}.

  Now, first we  take  this variant as the basic form and try to describe the electromagnetic form factors of nucleons with different PDF sets by quantitatively using the standard fitting procedure.
     Then we expand this form of $f^{q}(x)$ to a more complicated form which can have
     the parameters with the flavor dependence,
\ba
 f_{exp}^{q}(x)=
 2 \ \alpha_{H,E} \ z^{d}_{2}  [ \frac{(1-x)^{p_1 z^{d}_{1}}}{(x_0+x)^{p_2}}  ].
  \label{fqx}
\ea
As the result, the GPD functions will be
\ba
{\cal{H}}^{q} (x,t) \  = && \frac{2}{3} \ g^{u}(x)  \
     	 e^{ f^{u}_{exp} \ t}	
     	  -\frac{1}{3} \ g^{d}(x) \
        e^{ f^{d}_{exp} \ t },
\ea
\ba
{\cal{E}}^{q} (x,t)=&&\frac{k_{u}}{N_{u}}\frac{2}{3} \ g^{u}(x) (1-x)^{e_{u}}
     	 e^{ f_{exp}^{u} \ t}                            \nonumber \\
     	&&  +\frac{k_{d}}{N_{d}}\frac{1}{3} \ g^{d}(x)  \ (1-x)^{e_{d}} \
      e^{ f_{exp}^{d} \ t},
      \label{eud}
\ea
  with now $e_{u}$ and $e_{d}$ being the free fitting parameters.
According to  the normalization of the Sachs form factors, we calculate $N_d$ and $N_{u}$
 to obtain the anomalous magnetic moments of the quarks
$k_u=1.673  , \ \ \  k_d=-2.033 $.
  Here the parameters for the $d$ quark $z^{d}_{1}=1$ and $z^{d}_{2}=1$ if we take the
    flavor independent case and
  take as a free parameters in the contrary case.

\section{The sets of PDFs and experimental data of the nucleon form factors}

\begin{table}
 \caption{The sets of the PDFs with its basic parameters}
\label{Table-1}
\begin{center}
\begin{tabular}{|c|c|c|c|c|} \hline
N  & Model  & Reference & $g_{2}^{q}(x)$ & Order, ($Q_{0}^{2}$)          \\
     &         &               &                          \\ \hline
     &         &               &             &          \\
 1  &ABKM09    & \cite{ABKM09} & Eq. (\ref{ex2})& NNLO (9.)         \\
 2a & JR08a    & \cite{JR08}  & Eq. (\ref{sq2a})& NNLO (0.55)       \\
 2b & JR08b    & \cite{JR08}  & Eq. (\ref{sq2a})& NNLO (2.)      \\
 3  &ABM12    & \cite{ABM12} & Eq. (\ref{ex3})& NNLO (9.)       \\
 4a & KKT12a    & \cite{KKT12}  & Eq. (\ref{sqf})& NLO (4.)           \\
 4b & KKT12b    & \cite{KKT12}  & Eq. (\ref{sqf})& NLO (4.)          \\
 5a & GJR07d   & \cite{GJR07} & Eq. (\ref{sq2a})& LO (0.3)        \\
 5b & GJR07b   & \cite{GJR07} & Eq. (\ref{sq2a})&NLO (0.3)       \\
 5c & GJR07a   & \cite{GJR07} & Eq. (\ref{sq2a})&NLO (2.)         \\
 5d & GJR07c   & \cite{GJR07} & Eq. (\ref{sq2a})& NLO (0.3)        \\
 6a & MRST02  & \cite{MRST02}& Eq. (\ref{sq2a})& NLO (1.)         \\
 6b & MRST01  & \cite{MRST01}& Eq. (\ref{sq2a})& NLO (1.)             \\
 7a & GP08a    & \cite{GP08}  & Eq. (\ref{sq2b})& NLO (0.5)      \\
 7b & GP08b    & \cite{GP08}  & Eq. (\ref{sq2b})& NNLO (1.5)    \\
 7c & GP08c    & \cite{GP08}  & Eq. (\ref{sq2b})& NLO (2.)   \\
 7d & GP08d    & \cite{GP08}  & Eq. (\ref{sq2b})& NNLO (0.5)      \\
 8a & MRST09  & \cite{MRST09}& Eq. (\ref{sq2a})&  LO   (1.)       \\
 8b & MRST09  & \cite{MRST09}& Eq. (\ref{sq2a})&  NLO  (1.)        \\
 8c & MRST09  & \cite{MRST09}& Eq. (\ref{sq2a})& NNLO  (1.)         \\
 9  & MRST02P  & \cite{CTEQ6M}& Eq. (\ref{ex1})& NLO (1.3)       \\
10a & CJ12amin    & \cite{CJ12}  & Eq. (\ref{sq2a})& NLO(1.7)        \\
10b & CJ12am    & \cite{CJ12}  & Eq. (\ref{sq2a})& NLO(1.7)        \\
10c & CJ12bmid    & \cite{CJ12}  & Eq. (\ref{sq2a})& NLO(1.7)       \\
10c & CJ12cmax    & \cite{CJ12}  & Eq. (\ref{sq2a})& NLO(1.7)       \\
11  & MRSTR4  & \cite{MRST02}& Eq. (\ref{sq2a})&    NLO (1.3)        \\
   &         &              &                &         \\
 \hline
\end{tabular}
\end{center}
  \end{table}
\begin{table}
 \caption{Experimental data of the electromagnetic form factors)}
\label{Table-2}
\begin{center}
\begin{tabular}{|c|c|c|} \hline
 N points  &   Proton       &    References            \\
 111 & $G^{p}_{E} $ &  \cite{Andivadis94}; \cite{Walker94};  \cite{Boosted92}; \cite{Zhan11};
             \cite{Ron11};  \cite{Borkovsky75};  \cite{Arrington07};  \\
 196  & $G^{p}_{M}$  &  \cite{Andivadis94};   \cite{Walker94}; \cite{Sill93}; \cite{Boosted92}; \cite{Bartel73}; \cite{Grawford06}; \\
  & & \cite{Borkovsky75}; \cite{Arrington07};
    \\
87 & $\mu G^{p}_{E}/G^{p}_{M}  $  &  \cite{Walker94}; \cite{Bartel73}; \cite{Milbrath97};
   \cite{Jones06}; \cite{Gayou01}; \cite{Arrington07};   \\
    &              &             \\   \hline

   &   neutron      &                \\
13 & $G^{n}_{E} $ &  \cite{Bermuth03};   \cite{Glazier04}; \cite{Madey03};
             \cite{Rock82}; \cite{Eden94};  \cite{Becker99};  \cite{Zhu01};  \\
         & &    \cite{Warren03};
               \cite{Rohe99}; \\
38 & $G^{n}_{M}$  & \cite{Kubon01}; \cite{Brooks00}; \cite{Lachnet00}; \cite{Markowitz93};
  \cite{Bruins95}; \\
6 & $\mu G^{n}_{E}/G^{n}_{M}  $  &  \cite{Riordan10}; \cite{Madey03};  \\
   &              &             \\
 \hline
\end{tabular}
\end{center}
  \end{table}

  The  PDF sets of the different Collaborations (see Table 1) have the common form
  \ba
 x g^{q}(x) = N_{q} g^{q}_{1}(x) \ g^{q}_{2}(x) ,
\label{sq0}
\ea
  where the basic part $g^{q}_{1}(x)$ has the same form for  all the sets
\ba
  g^{q}_{1}(x) =  x^{a_{1}}(1-x)^{a_{2}},
\label{sq1}
\ea
 which give the rough presentation at small and large $x$.
 The second part $g_{2}(x)$ inputs some corrections to the basic form and  has
  different forms,
\ba
 g^{q}_{2}(x) = (1+ a_{3} \sqrt{x} + a_{4} x),
\label{sq2a}
\ea
 in  \cite{MRST01,MRST02,MRST09,CJ12}, \cite{JR08},
 and with the additional power of $x$  in \cite{GP08},
\ba
 g^{q}_{2}(x) = (1+ a_{3} \sqrt{x} + a_{4} x + a_{5} x^{1.5}),
\label{sq2b}
\ea
or with the free power of $x$  \cite{KKT12},
\ba
 g^{q}_{2}(x) = (1+ a_{3} x^{a_{5}} + a_{4} x ).
\label{sqf}
\ea
 Some more complicated form with the exponential dependence was used
 in  \cite{CTEQ6M},
\ba
  g^{q}_{2}(x) = e^{a_{3} x} \ (1 + x e^{a_{4} x})^{a_{5}},
\label{ex1}
\ea
   and in power form  in \cite{ABKM09},
\ba
  g^{q}_{2}(x) =  x^{a_{3} x + a_{4} x^{2} },
\label{ex2}
\ea
 and with slightly different form in \cite{ABM12},
 \ba
  g^{q}_{2}(x)  = x^{a_{3} x + a_{4} x^{2} + a_{5} x^{3}}.
\label{ex3}
\ea

    The PDF sets are determined from the inelastic processes in  some bounded region of $x$.
     However, to obtain the form factors,  we have to integrate  over $x$ in the whole range $0 \div 1$.
    Hence, the behavior of  PDFs, when $x \rightarrow 0$ or $x \rightarrow 1$,  can  impact  the
    form of the calculated form factors.

\section{Analysis and Results}

\begin{figure}
\includegraphics[width=.4\textwidth]{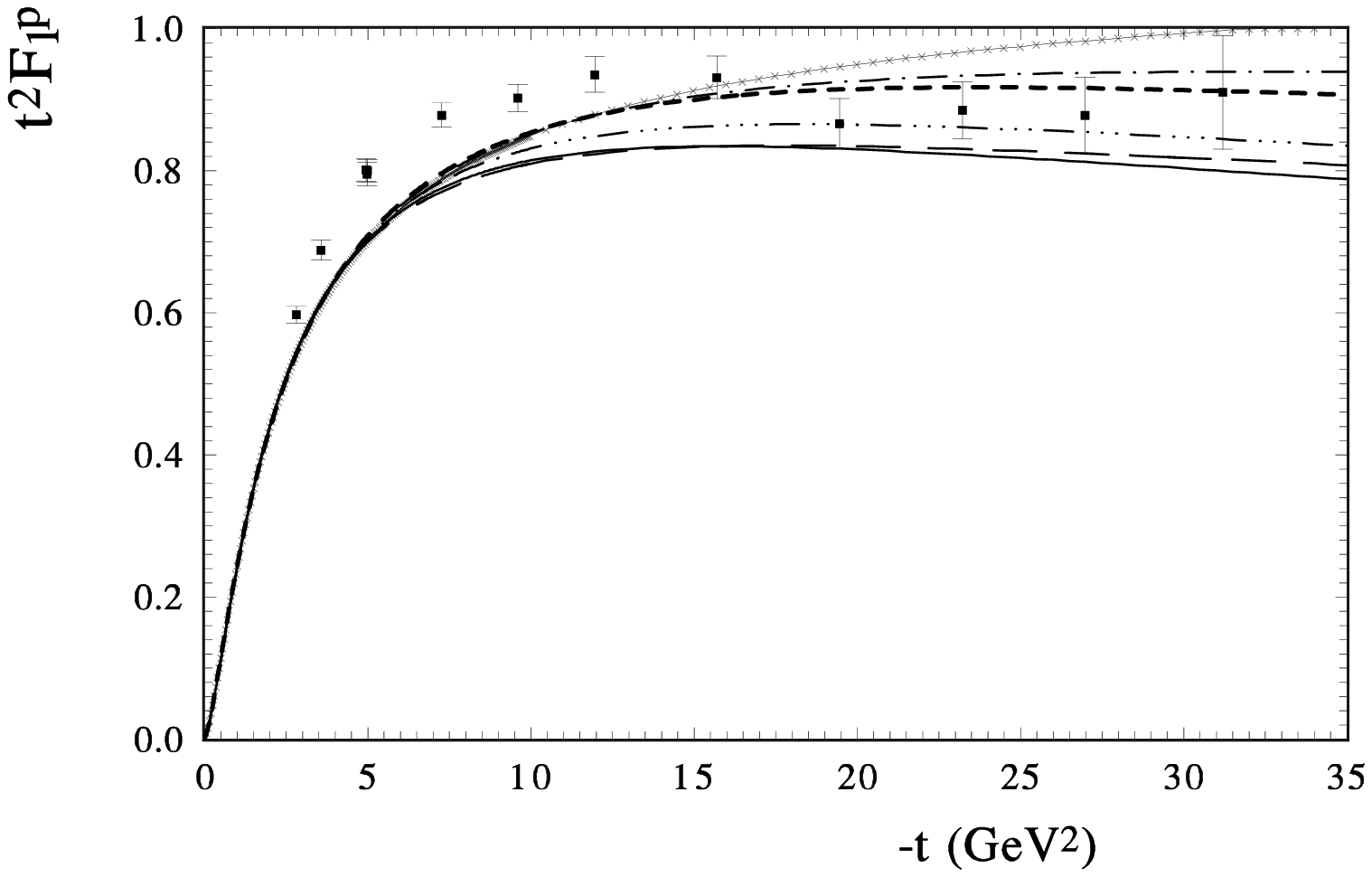} 
\includegraphics[width=.4\textwidth]{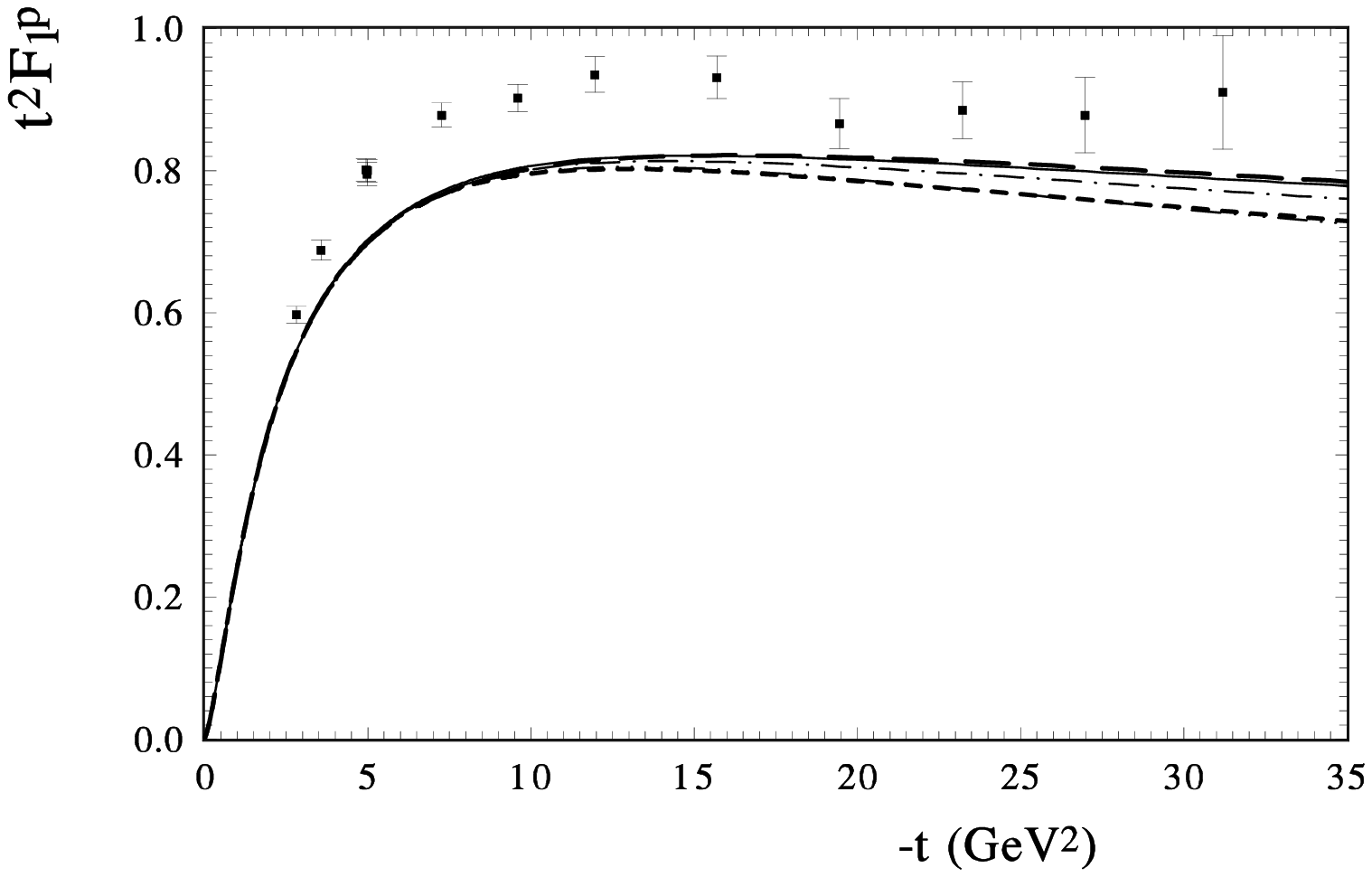} 
\caption{ Proton Dirac form factor multiplied by $t^2$
in (a) top panel, the basic variant I Eq.(\ref{fbv})
and (b) bottom panel, the variant IV  Eq.(\ref{fqx}).
  }
\label{Fig_1}
\end{figure}

\begin{table}
 \caption{The sum of $\chi^2$ for the different PDFs sets
  and with different number of the fitting parameters.}
\label{Table-3}
\begin{center}
\begin{tabular}{|c|c|c|c|c|c|c|c|} \hline
N& Model  & $\chi^{2}_{0}$ &$\chi^{2}_{+1p}$ & $\chi^{2}_{+1p}$ &  $\chi^{2}_{+2p}$ &   $\chi^{2}_{+3p}$  & $\chi^{2}_{+4p}$\\
   &           &         &        &        &       &       &            \\ \hline
 1 & ABKM09    &    984  & 984    &  953   &  936  &  903  & 872      \\
 2a& JR08a      &  $1119$ &   861  &  891   &  861  &  860  & 857      \\
 2b& JR08b      &  $1242$ &  1242  &  880   &  868  &  868  & 864       \\
 3 & ABM12     &  $1036$ & 1033   & 1031   & 1020  &  919  & 904      \\
 4a& KKT12a  8  &  $1170$ & 1133   & 1170   & 1108  &  934  & 888       \\
 4b& KKT12b     &  $1074$ & 1074   & 1064   & 1064  &  1036 & 988         \\
 5a& GJR07d    &  $1772$ & 1042   & 1553   &  936  &  884  & 878           \\
 5b& GJR07b    &  $1172$ & 1078   &  992   &  947  &  887  & 865            \\
 5c& GJR07a    &  $1215$ & 1214   & 1079   & 1024  &  940  & 894          \\
 5d& GJR07c    &  $8423$ & 1230   & 7279   & 1042  &  954  & 891            \\
 6a& MRST02       &  $1089$ & 1041   & 1035   & 1013  &  932  & 905              \\
 6b& MRST01       &  $1167$ & 1002   & 1129   &  999  &  898  & 873                \\
 7a& GP08a     &  $2189$ & 1575   & 1495   & 1017  &  886  & 879         \\
 7b& GP08b     &  $1423$ & 1382   & 1009   &  988  &  891  & 888          \\
 7c& GP08c     &  $1278$ & 1226   &  991   &  974  &  898  & 892           \\
 7d& GP08d     &  $4587$ & 2484   & 4575   & 3483  & 2388  &2388          \\
 8a& MRST09a      &  $1785$ & 1184   & 1598   & 1107  &  974  & 887               \\
 8b& MRST09b      &  $1382$ & 1226   & 1149   & 1052  &  972  & 894             \\
 8c& MRST09c      &  $1260$ & 1168   & 1005   &  960  &  930  & 881             \\
 9 & MR02P      &  $1344$ & 1187   & 1120   & 1044  &  946  & 875              \\
10a& O12a      &  $1523$ & 1458   &  1080  & 1054  &  1007 & 932             \\
10b& O12am     &  $1534$ & 1468   &  1077  & 1050  &  1007 & 932              \\
10c& O12b      &  $1377$ & 1361   &  1134  & 1127  &  1052 & 958                 \\
10d& O12c      &  $1366$ & 1359   &  1192  & 1191  &  1085 & 981          \\
11 & MRST02R4     &  $2360$ & 2358   & 1879   & 1819  & 1786  &1780              \\
   &       &      &       &       &      &       &   \\
 \hline
\end{tabular}
\end{center}
\end{table}

   We analyzed  the PDF sets in  five cases:  first,  with minimum free parameters and flavor independence
   $f(x,t)$ Eq.(\ref{fbv}) (basic variant), as was made in \cite{ST-PRDGPD},
        and then with an increase in the number of  free parameters
   (a)  free $p_{1}$ (both $u$ and $d$ quarks have the same power),
    (b) fixed $p_{1}$   and made as free $z_{1}$ (the $u$ quarks correspond to the basic variant and $d$  quark has the free power dependence),
   (c)  made  free $p_{1}$ and $z_{1}$ (both quarks have the independent power dependence),
   (d) using free $p_{1}$, $z_{1}$ and $z_{2}$ (the slopes of the $u$ and $d$ quarks can be different).
     The  last two variants already have a small difference in $\chi^2$ for  most variants of PDFs,
     as can be seen in  Table 3 ( Coulomb 6 and 7).
     So including  extra free parameters leads to small decreasing of $\chi^2$ and does not give
     new information about the properties of PDFs.  We research   also the case
     with the supplementary term of $x$ in  $f(x)$ in the form
     $z_{3} \ x \ (1-x)$. The results are shown in the last column of Table 3. We can see that this variant does not give additional useful information about the PDF sets.

     The PDF sets were taken  as $24$  variants in different works with taking into account the leading order (LO),
      next leading order (NLO) and next-next leading order {NNLO) in $\alpha_{s}$ of QCD (Table 1).
      The experimental data on the electromagnetic form factors were represented by
     $446$ experimental points.

\begin{figure}
\includegraphics[width=.4\textwidth]{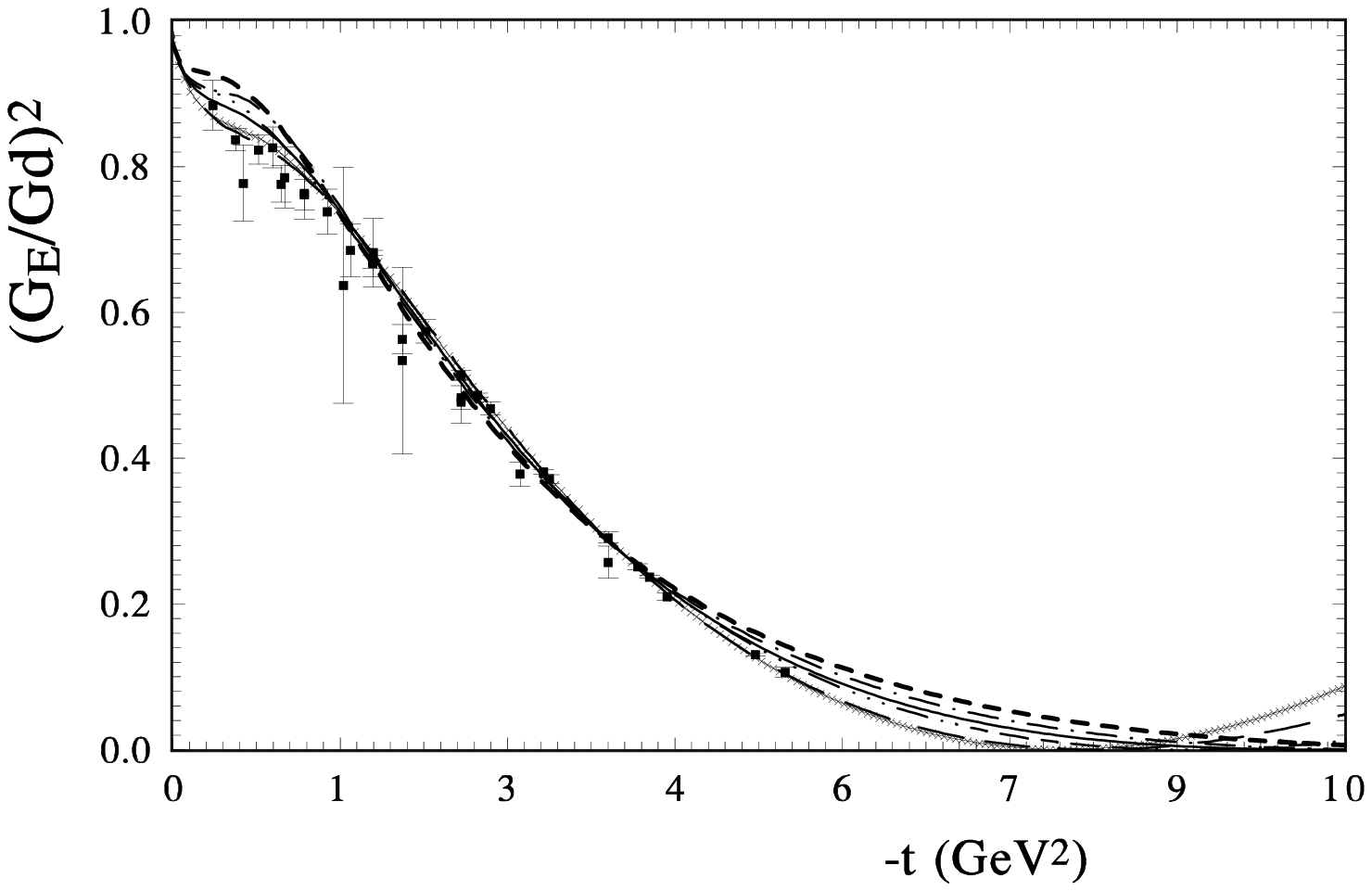} 
\includegraphics[width=.4\textwidth]{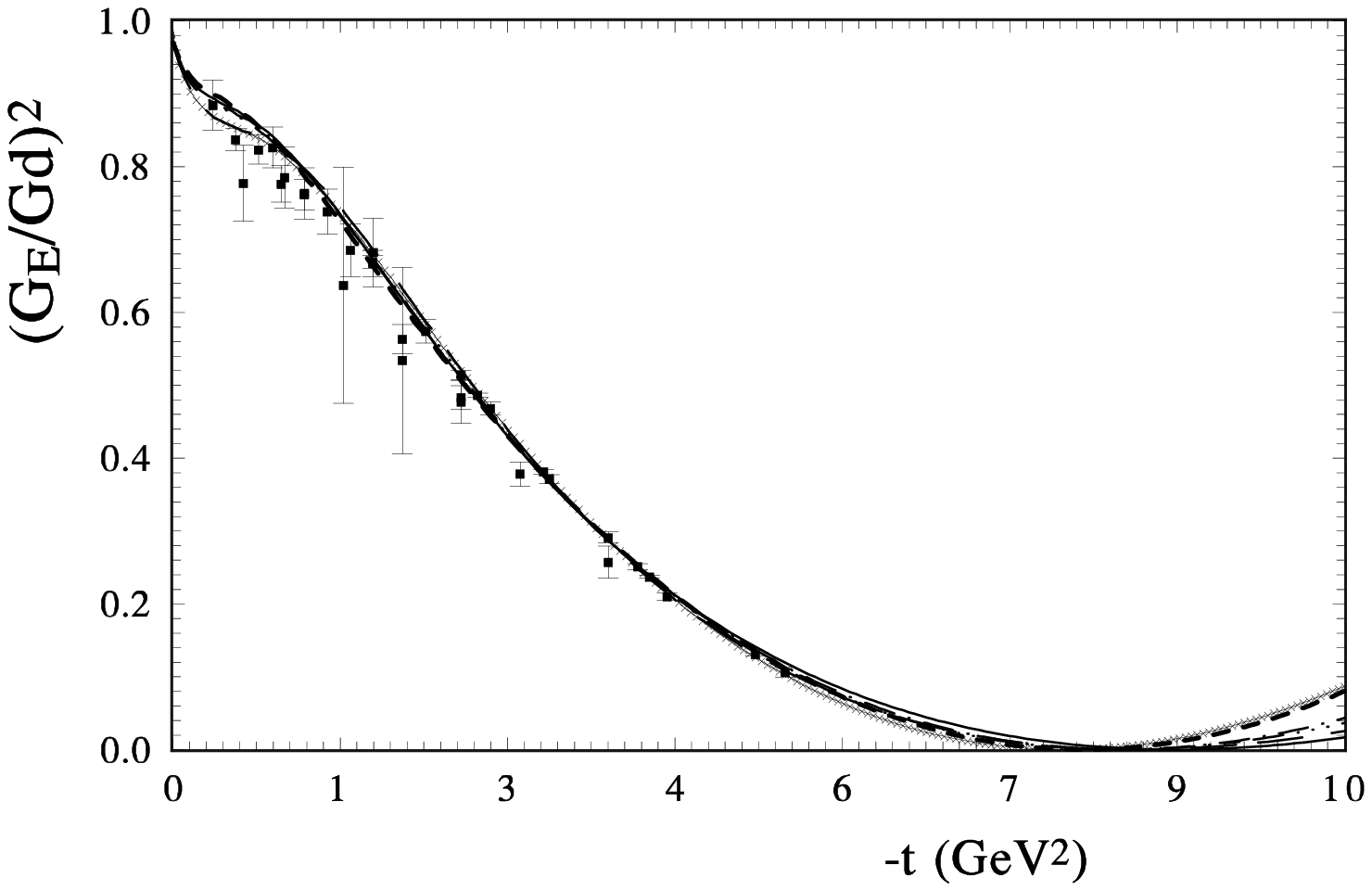} 
\caption{ Proton $(G_{E}/Gd)^2$ in
(a) top panel, the basic variant I Eq.(\ref{fbv}) and
(b) bottom panel, the variant IV  Eq.(\ref{fqx}).
 The  data for $F_1^{p}$ are from \cite{Sill93}.
  }
\label{Fig_2}
\end{figure}

   The whole sets of the experimental data are presented in Table 2.
      We include  both compilations of the experimental data \cite{Andivadis94} and \cite{Arrington07}.
      The sets of the data have  various corrections and the different methods taking into account the
      systematical errors.  So we take into account only the statistical errors.  Of course, we obtain sufficiently
      large $\sum \chi^{2}_{i}$. However, we are interested in the difference between
        $\chi^2$ obtained with the different PDF sets
      and the  number of  free parameters.

\begin{table}
 \caption{Basic parameters of the model with the different PDFs sets}
\label{Table-4}
\begin{center}
\begin{tabular}{|c|c|c||c|c|c|c|c|c|} \hline
N & $p_1$  & $p_2$         &  $\alpha_{H}$   & $\alpha_{E}$  & $x_{0}$ & $e_{u}$          & $e_{d}$         \\
    &  fixed & $ \pm 0.005$ & $ \pm 0.004$ &  $ \pm 0.003$  &  $ \pm 0.004$  &  $ \pm 0.05$ &  $ \pm 0.05$ \\ \hline
 1 & 2.0   & 0.507 & 0.377     & $0.382$  & $0.007$      & 2.69   & 0.09  \\
 2a& 2.0   & 0.382 & 0.641     &  $0.735$ &  $0.001$     & 1.47   &-0.53   \\
 2b& 2.0   & 0.428 & 0.487     &  $0.567$ &  $0.004$     & 1.69   &-0.55   \\
 3 & 2.0   & 0.510 & 0.377     & $0.370$  & $0.008$      & 2.86   & 0.24   \\
 4a& 2.0   & 0.433 & 0.491     &  $0.479$ &  $0.008$     & 2.26   &-0.04  \\
 4b& 2.0   & 0.422 & 0.495     &  $0.508$ &  $0.008$     & 1.97   &-0.04  \\
 5a& 2.0   & 0.238 & 0.849     &  $0.837$ &  $0.000$     & 1.87   & 0.12    \\
 5b& 2.0   & 0.342 & 0.683     &  $0.716$ &  $0.003$     & 1.79   &-0.09     \\
 5c& 2.0   & 0.415 & 0.498     &  $0.508$ &  $0.010$     & 2.29   & 0.09  \\
 5d& 2.0   & 0.140 & 0.974     &  $1.019$ &  $0.000$     & 1.49   &-0.23       \\
 6a& 2.0   & 0.421 & 0.565     &  $0.550$ &  $0.005$     & 2.24   & 0.21  \\
 6b& 2.0   & 0.392 & 0.596     &  $0.576$ &  $0.004$     & 2.20   & 0.21  \\
 7a& 2.0   & 0.328 & 0.694     &  $0.878$ &  $0.007$     & 1.13   &-0.99  \\
 7b& 2.0   & 0.355 & 0.525     &  $0.657$ &  $0.006$     & 1.21   &-1.12  \\
 7c& 2.0   & 0.315 & 0.547     &  $0.662$ &  $0.003$     & 1.27   &-0.97  \\
 7d& 2.0   & 0.449 & 0.763     &  $0.635$ &  $0.000$     & 1.60   & 0.59  \\
 8a& 2.0   & 0.218 & 0.776     &  $0.841$ &  $0.000$     & 1.46   &-0.38  \\
 8b& 2.0   & 0.326 & 0.632     &  $0.710$ &  $0.002$     & 1.54   &-0.43  \\
 8c& 2.0   & 0.357 & 0.600     &  $0.681$ &  $0.003$     & 1.56   &-0.46  \\
 9 & 2.0   & 0.389 & 0.528     &  $0.561$ &  $0.002$     & 2.05   &-0.11  \\
10a& 2.0   & 0.377 & 0.533     &  $0.615$ &  $0.001$     & 1.71   &-0.44  \\
10b& 2.0   & 0.378 & 0.533     &  $0.613$ &  $0.001$     & 1.73   &-0.44  \\
10c& 2.0   & 0.377 & 0.539     &  $0.628$ &  $0.000$     & 1.43   &-0.61  \\
10c& 2.0   & 0.384 & 0.536     &  $0.619$ &  $0.000$     & 1.41   &-0.60  \\
11 & 2.0   & 0.388 & 0.579     &  $0.610$ &  $0.002$     & 1.52fix   &0.31fix  \\
   &       &       &           &          &              &        &       \\
 \hline
\end{tabular}
\end{center}
  \end{table}

\begin{figure}
\includegraphics[width=.4\textwidth]{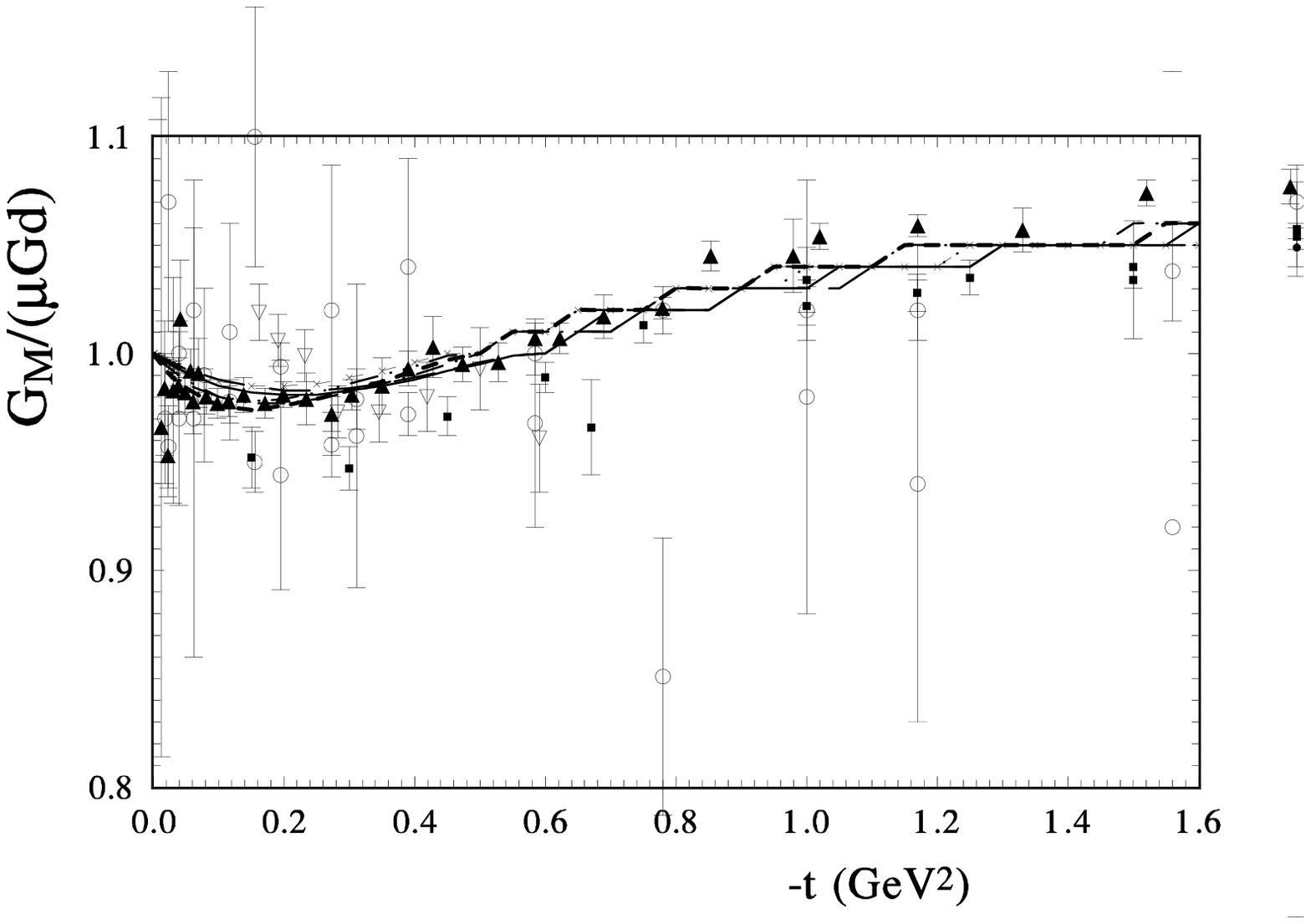} 
\includegraphics[width=.4\textwidth]{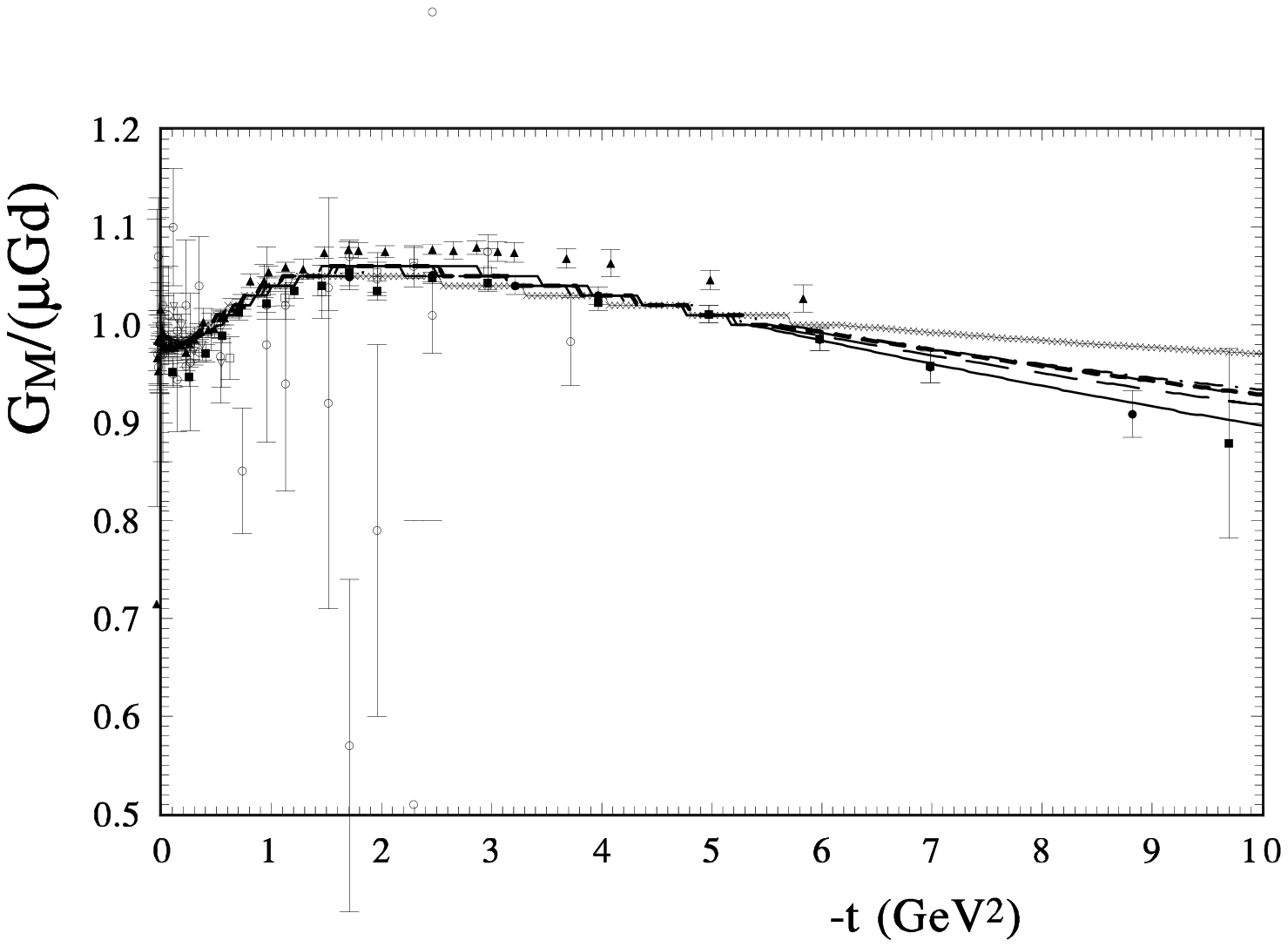} 
\includegraphics[width=.4\textwidth]{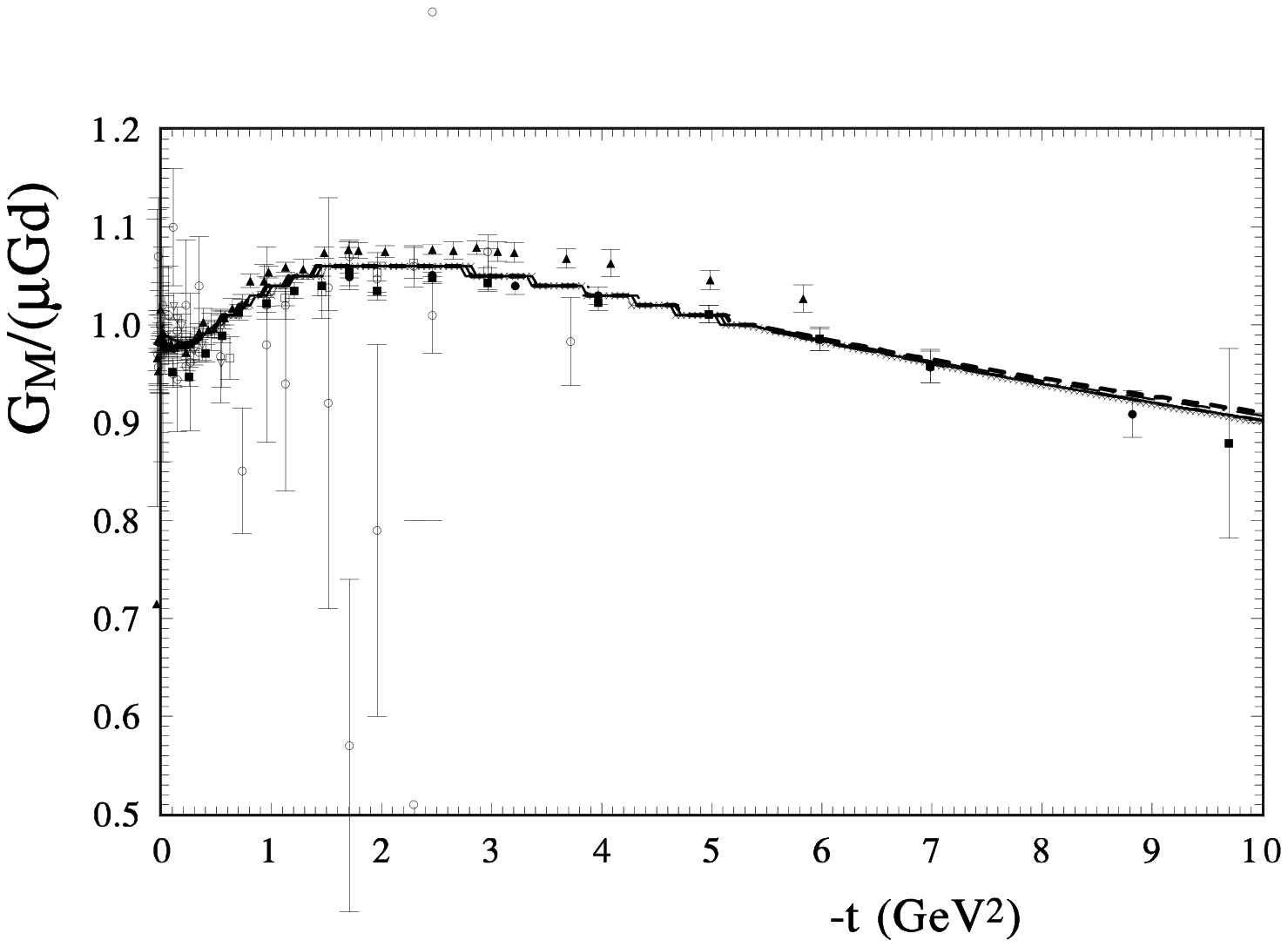} 
\caption{ Proton $G_{M}/(\mu Gd)$ in
(a) top and middle panels, the basic variant I Eq.(\ref{fbv}) and
(b) bottom panel, the variant IV  Eq.(\ref{fqx}).
  }
\label{Fig_3}
\end{figure}

\begin{table*}
 \caption{The fitting parameters of the GPDs with flavor dependence)}
\label{Table-3}
\begin{center}
\begin{tabular}{|c|c||c|c|c|c|c|c|c|c|c|} \hline
   &           &               &               &          &  & & & & & \\
N  &Model  & $p_1$ & $p_2$ &  $\alpha_{H}$ & $\alpha_{E}$  & $e_{u}$ & $e_{d}$ & $x_0$ & $z_{1}$ & $z_{2}$ \\
 & & $\pm0.02$ &  $\pm0.01$ &  $\pm0.01$& $\pm0.03$   &$\pm0.07$ & $\pm0.05$& $\pm0.002$& $\pm0.03$& $\pm0.03$\\ \hline
   &               &               &               &      &    & & & & &   \\
 1  & ABKM09 & 2.11  & 0.42 & $0.45$  &  $ 0.57$ &$0.67$   &-1.88 & 0.004 & $0.57$ & $ 0.91$    \\
 2a & JR08a  & 1.93 & 0.42 & 0.62  &  $0.75$  &$0.74$ &$-1.37$ & $0.006$ & $0.76$ & $0.98$  \\
 2b & JR08b  & 2.05 & 0.40 & 0.54  &  $0.67$&$0.71$ &$-1.68$ & $0.004$ & $ 0.71$ & $1.00 $  \\
 3  & ABM12  & 2.13 & 0.406 & 0.47  &  $0.61$ &$0.55$ &$-2.06$ & $0.002$ & $0.49$ & $0.87$  \\
 4a & GJR07d & 2.05 & 0.35 & 0.57  &  $0.74$   &$0.47$ &$-1.85$ & $0.006$ & $0.55$ & $0.89$  \\
 4b & GJR07b & 1.90 & 0.38 & 0.65  &  $0.80$   &$0.62$  &$-1.37$ & $0.009$ & $0.66$ & $0.91$  \\
 4c & GJR07a & 1.48 & 0.35 & 0.60  &  $0.78$   &$0.32$ &$-1.43$ & $0.007$ & $0.58$ & $0.91$  \\
 4d & GJR07c & 1.81 & 0.29 & 0.75  &  $0.89$   &$0.67$ &$-1.06$ & $0.002$ & $0.67$ & $0.90$  \\
 5a & Kh-12a & 1.97 & 0.40  & 0.52  &  $0.68$   &$0.44$ &$-2.00$ & $0.007$ & $0.51$ & $0.83$ \\
 5b & Kh-12b & 2.00 & 0.40 & 0.51  &  $0.58$   &$1.38$ &$-0.42$ & $0.005$ & $0.91$ & $ 0.96$  \\
 6a & MRST02 & 1.94 & 0.42   & 0.57  &  $0.68$  &$0.82$ &$-1.20$ & $0.006$ & $0.63$ & $ 0.88$  \\
 6b & MRST01 & 1.87 & 0.43   & 0.56  &  $0.68$   &$0.71$ &$-1.28$ & $0.01$ & $0.58$ & $ 0.86$  \\
 7a & GP08a  & 1.74 & 0.58 & 0.51  &  $0.53$   &$1.43$ &$-0.49$ & $0.04$ & $1.05$ & $1.21$  \\
 7b & GP08b  & 1.99 & 0.39 & 0.52  &  $0.56$ &$1.47$ &$-0.75$ & $0.008$ & $1.03$ & $1.17$  \\
 7c & GP08c  & 1.98 & 0.35 & 0.53  &  $0.57$ &$1.48$ &$-0.69$ & $0.005$ & $ 1.02$ & $ 1.13$  \\
 7d & GP08d  & 1.66 & 0.54 & 0.56  &  $0.73$ &$0.12$ &$-1.82$ & $0.00$ & $ 0.51$ & $ 0.71$  \\
 8a & MRST09A   & 1.80 & 0.28 & 0.67  &  $0.88$   &$0.34$  &$-1.70$ & $0.002$ & $0.64$ & $0.90$  \\
 8b & MRST09B   & 1.85 & 0.40 & 0.57  &  $0.75$   &$0.35$  &$-1.88$ & $0.009$ & $0.63$ & $0.92$  \\
 8c & MRST09C   & 1.89 & 0.41 & 0.57  &  $0.73$  &$0.46$ &$-1.77$ & $0.01$ & $0.66$ & $0.94$  \\
 9  & MR02P & 1.87 & 0.43   & 0.50  &  $0.65$  &$0.46$ &$-1.81$ & $0.008$ & $0.55$ & $ 0.90$  \\
10a & O12A  & 1.92 & 0.40 & 0.53  &  $0.71$   &$0.27$  &$-2.18$ & $0.003$ & $0.59$ & $0.94$  \\
10b & O12Am  & 1.92 & 0.40 & 0.53  &  $0.72$  &$0.27$ &$-2.19$ & $0.003$ & $0.59$ & $0.94$  \\
10c & O12C  & 1.94 & 0.39 & 0.54  &  $0.74$ &$0.26$ &$-2.26$ & $0.001$ & $ 0.63$ & $0.92$  \\
10c & O12D  & 1.97 & 0.37 & 0.55  &  $0.76$  &$0.26$ &$-2.26$ & $0.00$ & $0.64$ & $0.91$  \\
11  & MRST02R4 & 1.88 & 0.48   & 0.51  &  $0.51$    &$1.52$ & $0.31fix$& $0.001$ & $0.86$  & $0.97$  \\
    &              &               &               &          &  & & & & & \\
 \hline
\end{tabular}
\end{center}
  \end{table*}

      In the final variant,  most of the PDF sets gave  approximately the same $\chi^2$
      (Table 3).
      On this background of  PDFs,  one variant of {\bf GP08d}  \cite{GP08} and  all variants
      of {\bf O12} \cite{CJ12} are essentially different and have large $\chi^{2}$.
       In the last row of  Table 3, we show the calculation of the {\bf MRST02} \cite{MRST02}
      with the fixed  parameters used by \cite{R04} and by us \cite{ST-PRDGPD}.
      In this case $\chi^2$, is two times larger, but, on the whole, it  confirms  our qualitative model.
       The best descriptions were obtained with the PDF sets {\bf ABKM09} \cite{ABKM09}
       and   {\bf JR08} \cite{JR08}.
      In this case, all 6 variants of the $t$ dependence gave a very close size of  $\chi^2$.
       Also, we obtained a good description with the PDF sets  {\bf ABM12}  \cite{ABM12} and {\bf KKT12} \cite{KKT12}.
        It is interesting to note that the good result was obtained with the sufficiently  old PDF sets {\bf MRST02}  \cite{MRST02} and {\bf MRST01} \cite{MRST01}.

       In most part, the best descriptions of the electromagnetic nucleon form factors
       were given by PDFs with the non-power forms of $g_{1}^{q}(x)$ eqs.(\ref{ex1})-(\ref{ex3}). However, the PDFs {\bf JR08}
       used the standard form of the $g_{1}^{q}(x)$ though with free power of $x$, Eq.(\ref{sqf}),
        instead of the standard $\sqrt{x}$.


\begin{figure}
\includegraphics[width=.4\textwidth]{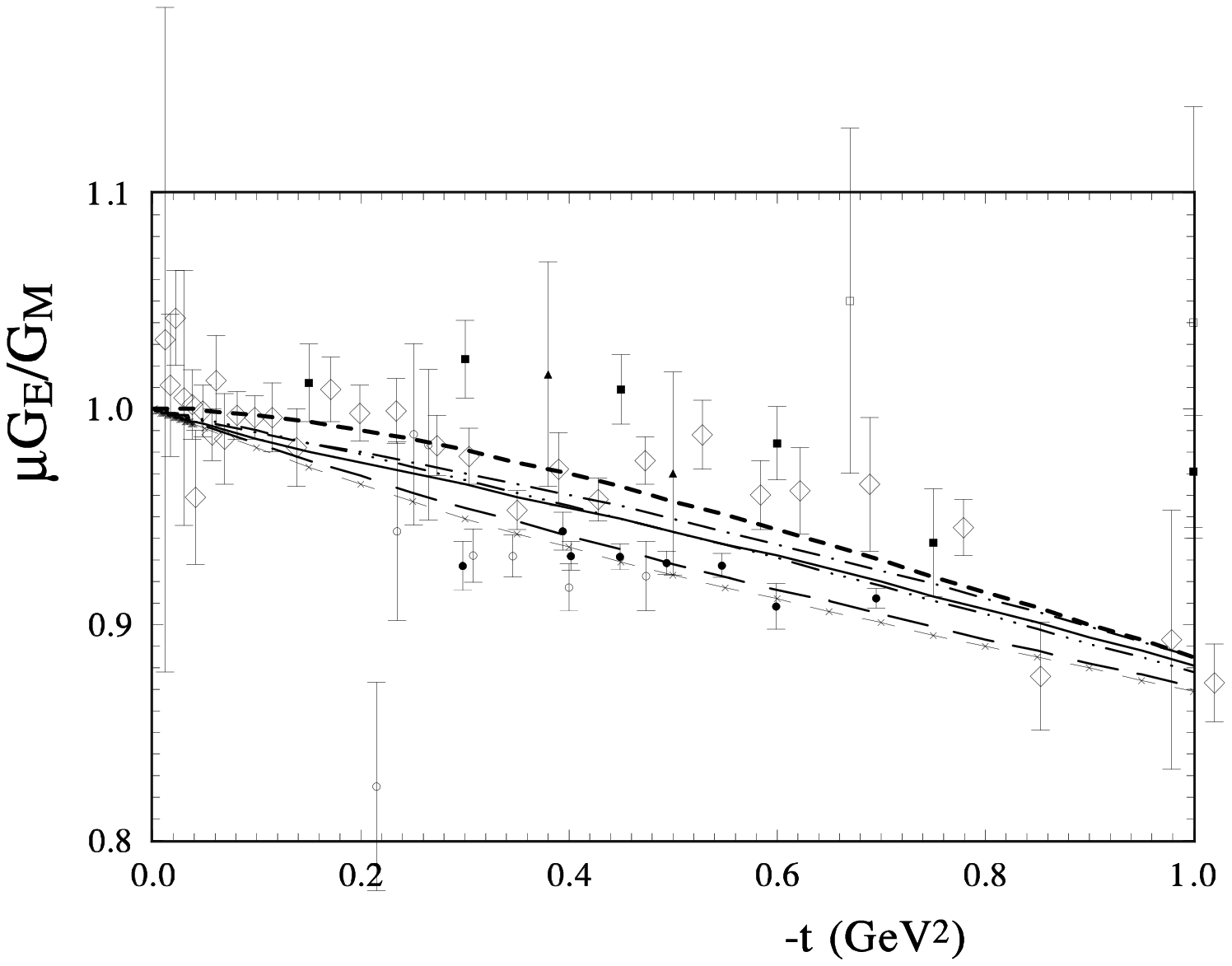} 
\includegraphics[width=.4\textwidth]{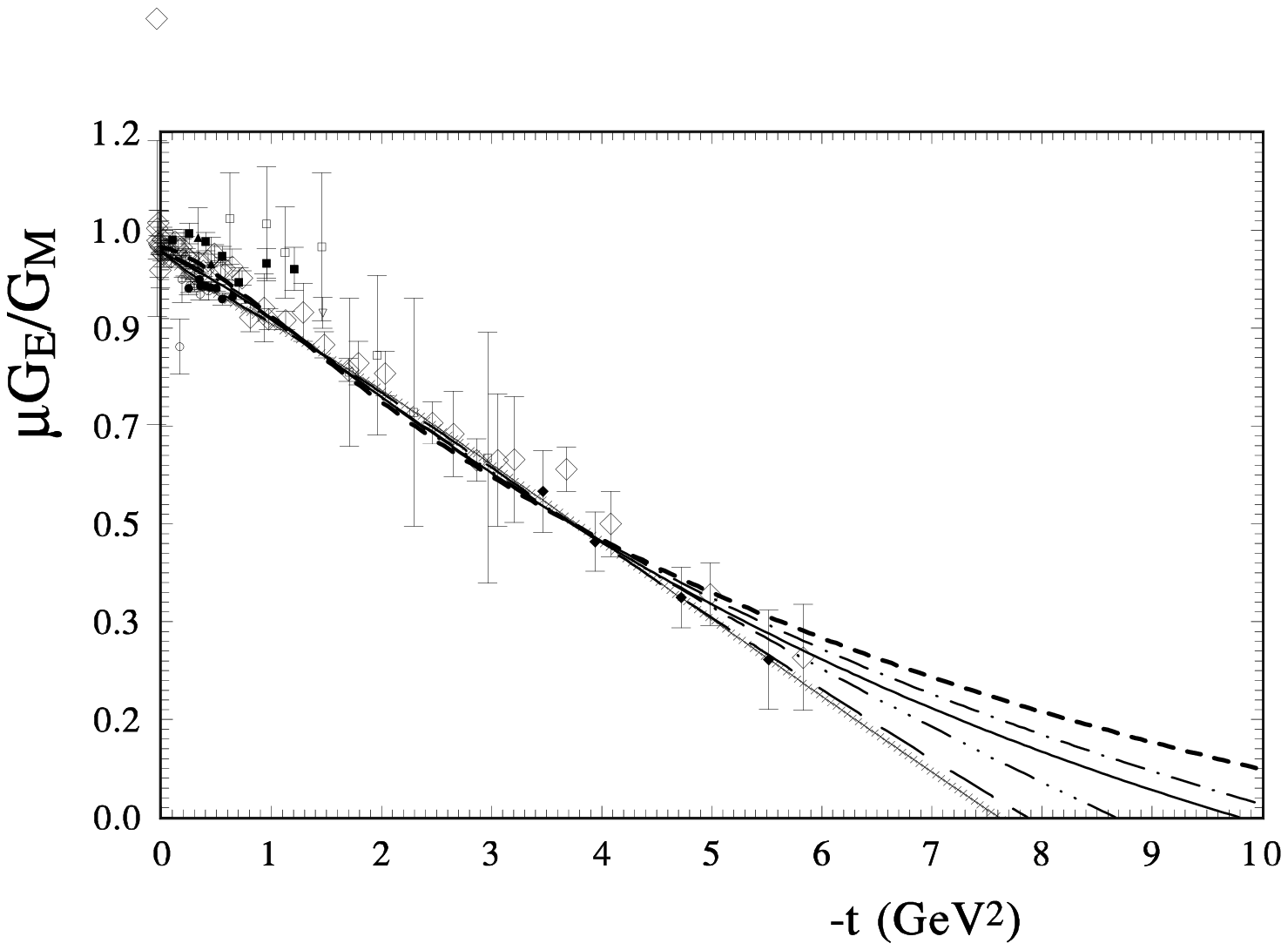} 
\includegraphics[width=.4\textwidth]{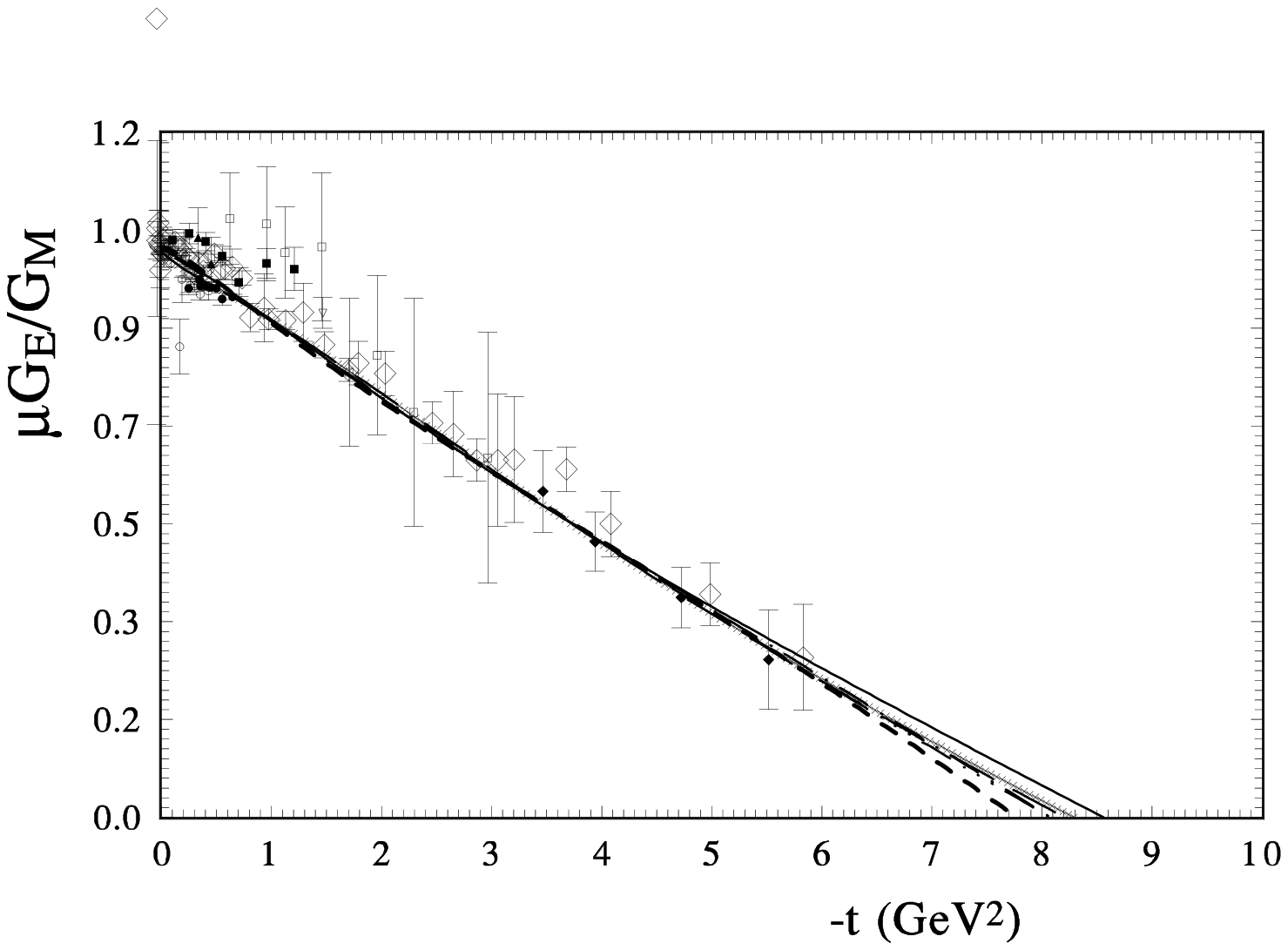} 
\caption{ Proton $\mu G_{E}/G_{M}$ in
(a) top and middle panels, the basic variant I Eq.(\ref{fbv}) and
(b) bottom panel, the variant IV  Eq.(\ref{fqx}).
  }
\label{Fig_4}
\end{figure}

      The impact of the difference forms of PDFs will be seen, maybe, in the description of the separate form factors.
        It is worth  noting, that the different PDF sets gave the similar descriptions in the proton form factors
   and a large difference in the description of the neutron form factors (see Fig.2, Fig.3, Fig.4 for the proton case
   and Fig. 6 , Fig. 7 for the neutron case).
   Probably, just the neutron data, in most part, lead to an essentially better description of the polarization data on the
   electromagnetic form factors.

    In our qualitative model we showed that the descriptions
   of the experimental data, related with the Rosenbluth and polarization methods,
    can be obtained
   by changing  the slopes of the $t$ dependence of the $u$ and $d$ quarks.
   In the present analysis all PDF sets led to the polarization case. Some difference was  obtained only at large   momentum transfer.

      In Table 4, the values of the parameters  of the basic variant are presented. Except some separate PDF sets,
      the  slopes of $H$ and $E$ have the mean value $0.55$ and $0.6$, respectively.
        As shown in our previous work
      \cite{ST-PRDGPD}, it is related with the Polarization variant of the obtained form factors. The Rosenbluth
      variant requires a large difference in these slopes. The value of the power $x$ in $f_{q}(x)$
      equals approximately $0.4$. We can see that the difference of  PDFs incoming in $E$ is distinguished, in most part,
      by the form of $u$-quark. It has the additional factor $(1-x)^{e_{u}}$ with $e_{u} \sim 2 $.
      Some PDF sets gave the large $\chi^2$, especially  one variant of {\bf GJR07} \cite{GJR07} and one variant of {\bf GP08} \cite{GP08}.
      If in last case we think it is the result of the Log-Log approximation; in the case of  \cite{GJR07}
      it is, maybe, the result of some misprint of the printed parameters.  

       The number of the parameters of the variant IV (with 4 additional free parameters) is
       given   in  Table 5. If in the previous case the power of $(1-x)$ in $f(x)$ was
       fixed by $p_{1}=2$, now its value does not go out far. The arithmetic mean value over all 24 variants of PDFs
       $\bar{p}_{1}=1.91$. In the best variants it is slightly above  $2$. In some other case it is less
       but, very likely, it reflects some attempt to improve the $x$ dependence of PDFs.
       The power of $x$ has arithmetic mean value $\bar{p}_2=0.39$. It coincides with the value in the previous  (basic) case.  The arithmetic mean of the slopes of $H$ and $E$ is $0.58$ and $0.72$.
       It is slightly above the previous case but again they do not strongly differ from each other.
       The large difference between variants I and IV comes from  $e_{u}$ and
       $e_{d}$. Now  $e_{u}$ decreases essentially and $e_{d}$ increases in absolute value.
        The coefficient $z_{1}$, reflecting the flavor dependence of the power $x$,
          differs  from unity. It is related with the exchange value of $e_{u}$ and $e_{d}$. However, the next flavor dependence
        $z_{2}$ , which reflects the flavor dependence of the slopes GPDs,  rest,  on the  average, near  unity. It is interesting that in the last variant {\bf Mrst02R4} with fixed $e_{u}$ and $ e_{d}$
        we obtained the values of both parameters $z_1$ and $z_2$ near unity.

     In Fig.1, it can be seen  that the basic variant with minimum  free parameters
   leads to a better description of the $t$ dependence of the data of the Dirac form factor $F_{1}(t)$.
    In this case,  PDFs {\bf CJ12a}, which gave one of the worst $\chi^2$ in the descriptions of all experimental data, gave the best description of $F_1^{p}(t)$.
   Note that the data on Fig.1 are related to the Rosenbluth method.
   Hence, it is very likely that these data are in contradiction with other data.

   The description of the electric form factor $G_{E}(t)$ is good in both variants (the basic (I) and with 4 additional free parameters (IV)) (see Fig.2). In this figure, we can see that the difference between
    PDFs  occurs  only in  the region of $t \approx 0.5 $ GeV$^2$ and $-t \sim 6 \div 7 $  GeV$^2$.
       The description of the magnetic form factor $G_{M}(t)$ is good in  all variants, especially
       with  4 additional parameters in the whole region of  momentum transfer (Fig. 3c). The basic variant also gave a good description at small $t$ ( Fig. 3a) and not a large difference at large $t$ (Fig. 3b).  Note that the best PDF {\bf ABKM09} (the low curve of Fig.3b) gave the maximum slope of $G_{M}^{p}$ and {\bf GP08} PDFs gave the minimal slope (upper curve of Fig.3b).
              As the result, we can see that the ratio $R_{p}= \mu G{E}(t)/G_{M}(t)$
     for the proton describes well all existing polarization data. Some difference occurs  at small $t$ and $-t > 6$  GeV$^2$ for the basic variant (I).
     Such  a difference practically disappears for the (IV) variant (see Fig.4c).  Note that the PDFs {\bf ABKM09} in the gave the medium result at
     small and large $t$ (Fig. 4a and Fig. 4b). The PDFs {\bf GP08} gave the minimal result at small and large $t$ and the maximal value was given the PDFs {\bf MRST09}.
     In Fig. 5, we show the difference between variant I and IV for the ratio $R_{p}$ for the different PDFs. It confirms our $\chi^2$ results. It can be seen that the difference is small up to $-t=4$ GeV$^2$ for  all PDFs, especially for {\bf ABKM09} and {\bf MRST02}. At large $t$ the difference grows fast especially for the PDF {\bf GP08} (upper curve on Fig. 5b) and PDFs {\bf O12} (low curve on Fig.5b).

      For the neutron form factors, which were obtained with the same parameters as for the proton case using  the isotopic symmetry, we obtained a larger difference for the PDF sets. Farther, we will show only the results for  variant IV (with four additional free parameters).
     It should be noted that the experimental data for neutron form factors are obtained, in most part,
     from the deuteron or Helium target. It may lead to an increase in the uncertanty at large $t$, as we
     do not know exactly the wave functions of the light nuclei at large $t$.

     The electric form factor of the  neutron $G_{E}^{n}(t)$ describes well  all PDFs, except {\bf GP08L} and {\bf GP08c} (upper curves on Fig. 6 ).
      At small $t$ the minimal values were given by    PDFs {\bf O12C} and maximal values   PDFs {\bf ABKM09} which gave the medium value at large $t$. The magnetic form factor $G_{M}^{n}(t)$ has a larger
         difference for  PDF (Fig.7). The large value is obtained with
          PDF  {\bf GP08a} and {\bf GP08c} and minimal values with    PDFs {\bf O12a} and {\bf O12c}.
          Hence, the ratio $R_{n}(t) = \mu_{n} G_{E}^{n}/G_{M}^{n}$ has a large difference already after $-t > 2 $ GeV$^{2}$ (Fig.8). The upper curves present the calculations with   {\bf GP08a} and {\bf GP08c}.
           The lower curves correspond to the calculations with the   PDFs {\bf O12c}.
           As usual, in most part, the calculations with PDFs {\bf ABKM09} are in the mid-position.
           We see that the slope of the ratio $R_{n}(t)$  decreases at large $t$ for most PDF sets.

           In Fig. 9, the ratio $R_{M}^{pn}(t) = \mu_{p} G_{M}^{n}(t)/ (\mu_{n} G_{M}^{p}(t))$ is given.
            The ratio has a small difference for different PDFs up to $-t=2$ GeV$^{2}$ and then this difference grows.
           The decreasing  ratio with $t$ is less for the PDFs {\bf GP08a} and {\bf GP08c} and larger
           for PDFs {\bf O12a} and {\bf O12c}.

           The $t$ dependence of the Dirac and Pauli form factors of the proton and neutron are shown in Fig. 10.
           The Dirac form factor has the same slope at large $t$ for the proton and neutron cases and for different PDF sets (Fig.10a).
           All PDF sets lead to  approximately the  same $t$ dependence for the proton Pauli form factor up to large $ t \sim 15 $ GeV$^{2}$. The neutron Pauli form factors decrease slightly faster
          and have a wider region for different PDF sets. The faster decreasing
          is due to PDFs {\bf O12A} and {\bf O12C},
          and low decreasing is given by  PDFs {\bf MRST02}.

\begin{figure}
\includegraphics[width=.4\textwidth]{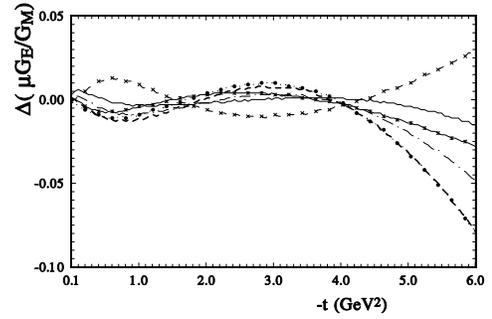} 
\caption{Difference between the basic variant I Eq.(\ref{fbv})
and variant IV Eq.(\ref{fqx}) of the ratio $R$.
  }
\label{Fig_5}
\end{figure}


\begin{figure}
\includegraphics[width=.4\textwidth]{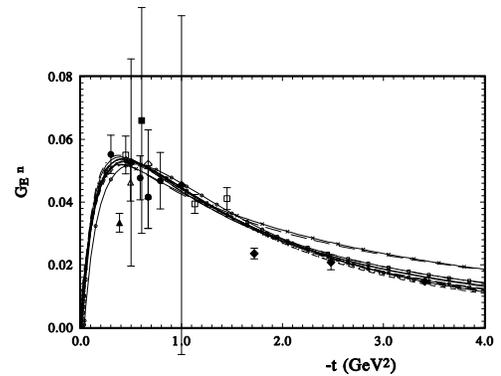} 
\caption{ Neutron $G_{E}^{n}(t)$.
  }
\label{Fig_6}
\end{figure}


\begin{figure}
\includegraphics[width=.4\textwidth]{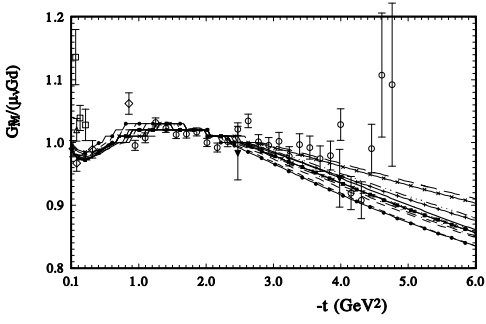} 
\caption{  Neutron $G_{M}^{n}/(\mu_{n} Gd)$.
  }
\label{Fig_7}
\end{figure}


\begin{figure}
\includegraphics[width=.4\textwidth]{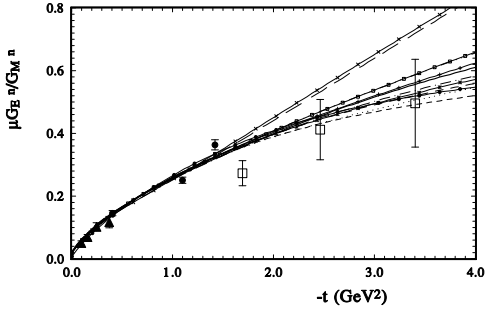} 
\caption{ Neutron $R_{n}(t)=\mu_{n} G_{E}^{n}(t) /G_{M}^{n}(t)$.
  }
\label{Fig_8}
\end{figure}


\begin{figure}
\includegraphics[width=.4\textwidth]{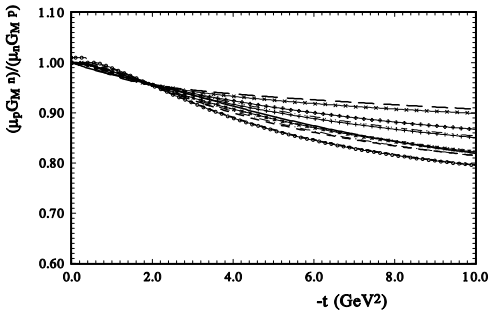} 
\caption{ Ratio $R_{M}^{pn}=\mu_{p}G_{M}^{n}/(\mu_{n}G_{M}^{p})$.
  }
\label{Fig_9}
\end{figure}

\begin{figure}
\includegraphics[width=.4\textwidth]{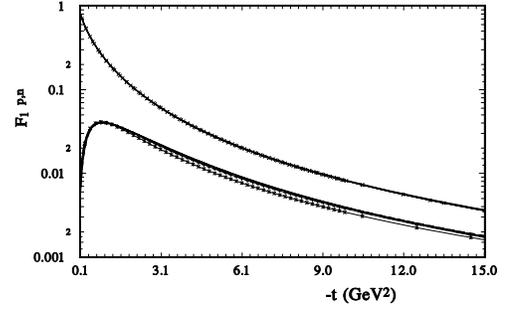} 
\includegraphics[width=.4\textwidth]{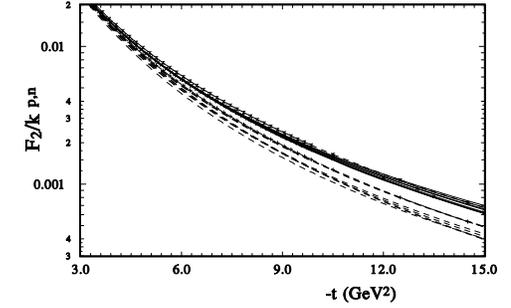} 
\caption{a) (top panel) Proton and neutron Dirac    form factors $F^{p,n}_{1}$;
b) (bottom panel)   Proton and neutron Pauly form factors $F^{p,n}_{2}/k_{p,n}$.
  }
\label{Fig_10}
\end{figure}

\section{Flavor dependence of GPDs }

   Let us examine  separate contributions of the $u$ and $d$ quarks to the
   electromagnetic form factors in our model of the $t$-dependence of  GPDs.
     In the basic variant I, all flavor dependence comes only from the difference of the coefficients
     $e_{u}$ and $e_{d}$, Eq.(\ref{eud}), in  PDFs incoming in $E(x,t)$. The coefficient $e_{d}$ is small and changing near zero for
     most PDFs. In these cases PDFs sets used non-power forms of $g_{2}^{q}(x)$.
     The PDFs, which used the standard Eq.(\ref{sq2a}) and Eq.(\ref{sq2b}),  have the large negative size of $e_{d}$
     and lead to the large $\chi^2$. The coefficient $e_{u}$ in this case is positive and large $1.5 < e_{u} < 2.5$.
      Hence, the $d$-distribution in $E(x,t)$ is, in most part, approximately the same as
     the $d$-distribution in $H(x,t)$. In the case of the additional free parameters (case IV, Table 5),
     the the coefficient $e_{d}$ increases up to $-2$ but the coefficient $e_{u}$ decreases and has  positive values.
     In this case, we include the parameters which take into account the flavor difference $z^{d}_{1}$ and $z^{d}_{2}$ of  GPDs, Eq.(\ref{fqx}). The value of  $z^{d}_{1}$ changes the behavior of the $d$ quark  $(1-x)^{p_{1} z^{d}_{1} }$.
     So its size heavily depends on the $x$-dependence of PDFs. However, the difference in the slope of the
     $u$ and $d$ quarks is small. The coefficient $z^{d}_{2}$ near $1 \pm 0.1$ is mostly of PDFs.
    It is very likely that the change of  the coefficient  reflects the problems of  minimization of  $\chi^2$   only.

\begin{figure}
\begin{center}
\includegraphics[width=0.4\textwidth] {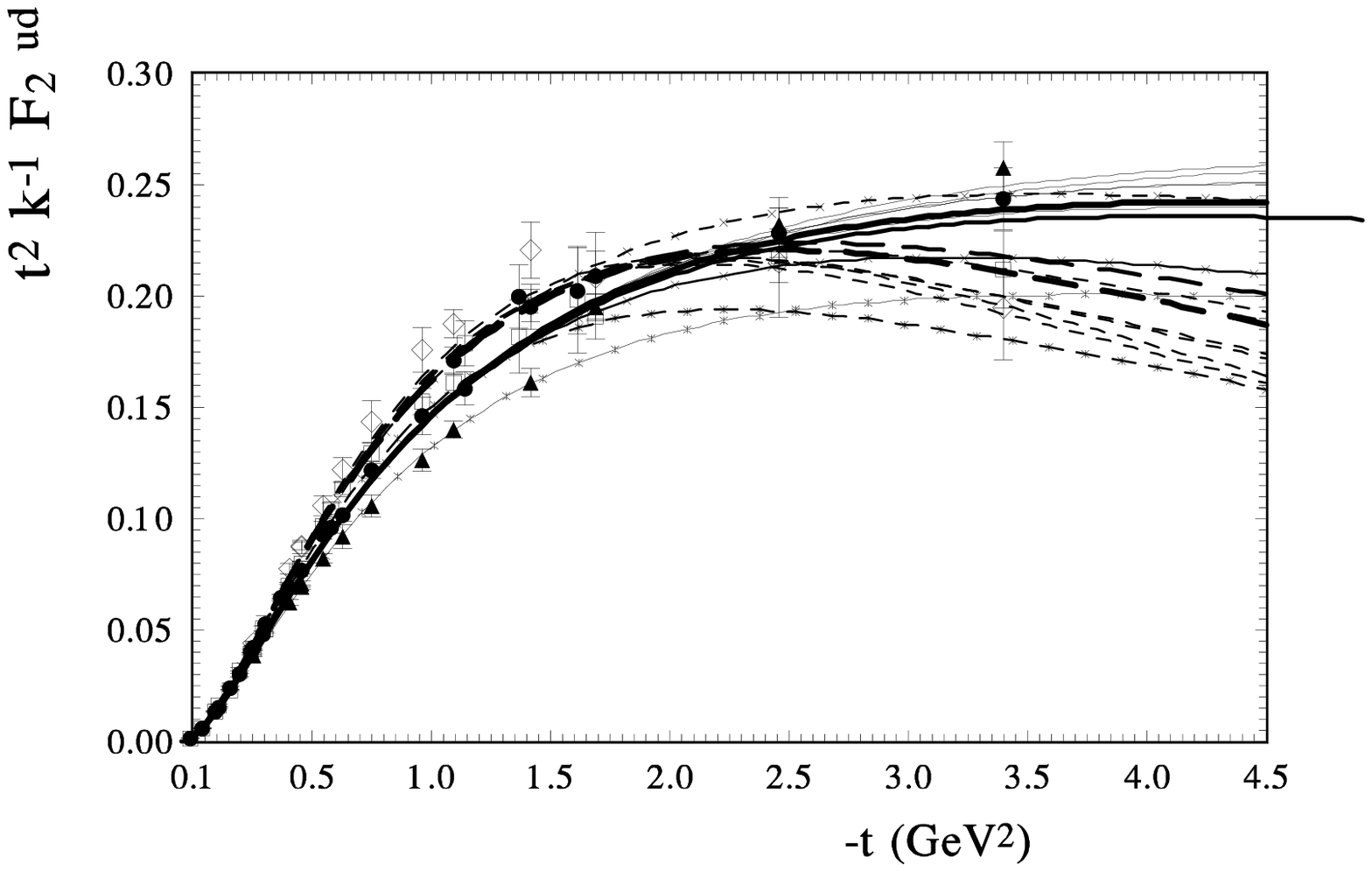}
\includegraphics[width=0.4\textwidth] {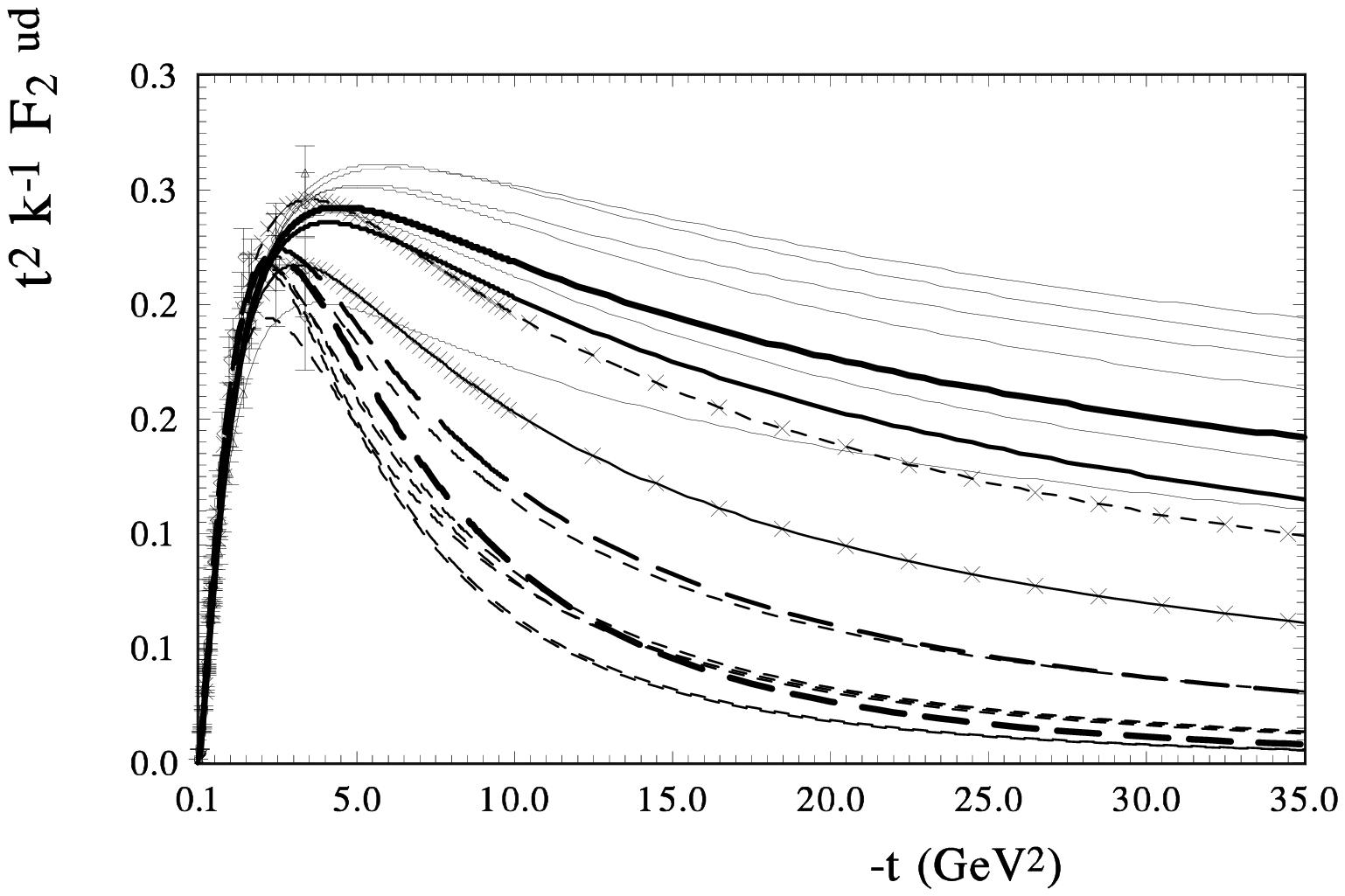}
\end{center}
 \caption{The $t$ dependence of $t^{2}  F^{u,d}_{2} / k_{u,d}$ of the $u$ and $d$ quarks contributions at small $t$  (top) and  at large $t$ (bottom panel).
 } \label{Fig_11}
\end{figure}

     In Fig.11, the obtained  $t$ dependence of $t^{2} \mu_{u,d} F^{u,d}_{2}(t)$ of the $u$ and $d$ quark contributions
      to the form factors at small $t$  (Fig. 11a) and  at large $t$ (Fig.11b) are presented.
     The dashed lines on these figures  reproduce the $d$ quark contribution and the hard lines reproduce
     the $u$ quark contribution.
      The contribution of the $d$ quark exceeds
     the contribution of the $u$ quark up to $-t= 2.5$ GeV$^{2}$  ( Fig. 11a).
      At larger momentum transfer the contribution of the $u$ quark
      exceeds the contribution of the $d$ quark, except the two cases of  PDFs.
       First, an essentially different picture is given by  PDFs {\bf GP08a}. In this case, the  contribution of the $d$ quark
      exceeds the contribution of the $u$ quark in the whole region of momentum transfer
      (upper dashed line with mark (x) for the $d$ quark  and the  low hard line with marks (x) for the $u$ quark in Fig. 11).
        For  PDFs {\bf MRST02} the contribution of the $u$ quark
      has the minimum value, compared with others, at $-t=2 $  GeV$^{2}$ and exceeds
       the $d$ quark contribution only
      after  $-t = 20 $  GeV$^{2}$.
       The minimum $d$ contribution is obtained with PDFs  {\bf O12} and {\bf MRST09a} (low dashed curves in Fig. 11b ).
        Close to these cases   PDFs  {\bf ABKM09} give the  $d$ contribution (thick long dashed curve in Fig. 11b).
        In  all cases, we see the same behavior of the $u$ and $d$ quark contribution at large momentum transfer.
        The slopes of all curves are practically the same.


\begin{figure}
\begin{center}
\includegraphics[width=0.4\textwidth] {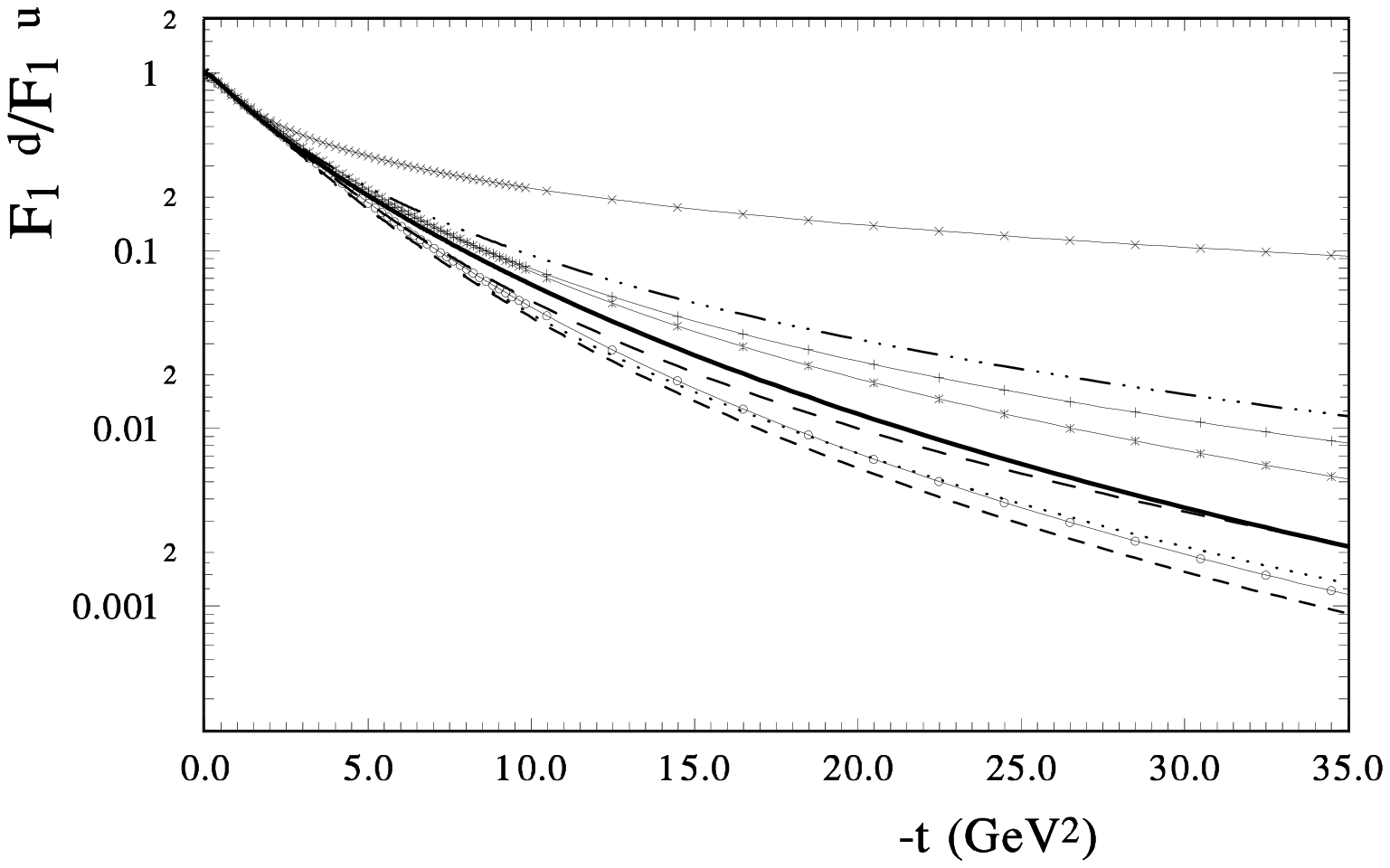}
\includegraphics[width=0.4\textwidth] {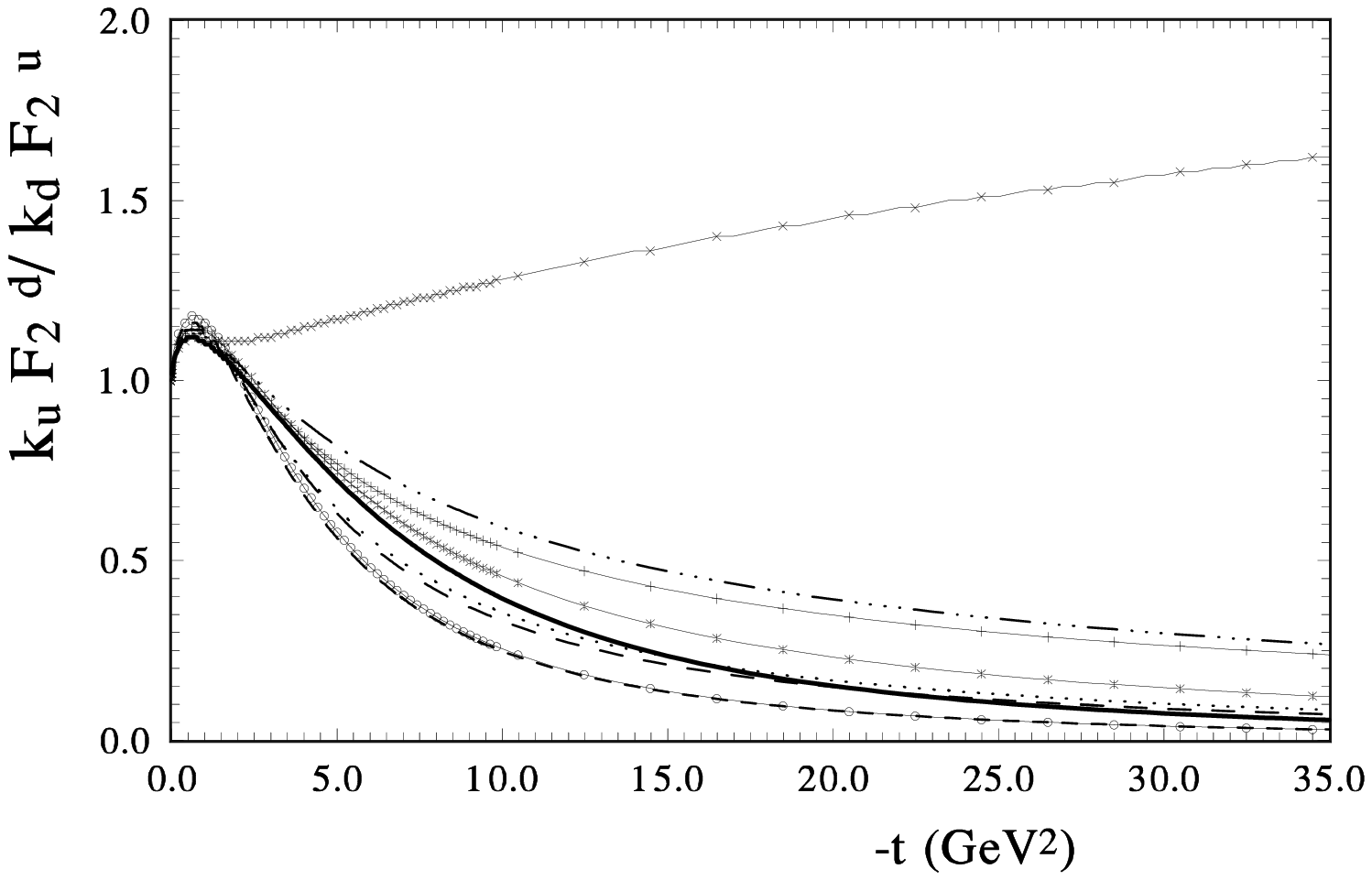}
\end{center}
 \caption{The ratio of the  $u$ and $d$ quarks form factors $F_{1}(t)$ (top) and  $F_{2}(t)$ (bottom panel)
  at large momentum transfer.
 } \label{Fig_12}
\end{figure}

   We obtain a remarkable picture  for the ratio of the contributions of the $u$ and $d$ quarks to
         Dirac and  Pauli form factors (Fig. 12). Again, we see a very different behavior for  PDFs  {\bf GP08a}
         (upper lines in Fig. 12a and 12b). Other PDFs give a similar behavior.
         The PDFs  {\bf O12} and {\bf MRST09a} (low dashed curves in Fig. 12a,b) give the  fastest decrease, and the PDFs  {\bf JR08} and {\bf GJR07} less decrease in the ratio of the $d$ and $u$ quarks.
        It is interesting that this  ratio of the contributions of the $u$ and $d$ quarks
        to the  Dirac and Pauli form factors has the same relative behavior of the different PDFs.
        The order of the curves practically repeats the Dirac and Pauli form factors.
        Of course,  the ratio for the Pauli form factor  less decreases at large
        momentum transfer than the ratio for the Dirac form factor.




\section{Gravitational form factors }

   Taking the
 matrix elements of energy-momentum  tensor $T_{\mu \nu}$
 instead of
   the electromagnetic current $J^{\mu}$
    \cite{Pagels,Ji97,PolyakovEMT}
 \ba
 \left\langle p^{\prime}|\hat{T}^{Q,G}_{\mu \nu} (0)|p\right\rangle &=& \bar{u}(p^{\prime}) \biggl[
 A^{Q,G}(t) \frac{\gamma_{mu}P_{\nu} }{2}   \\
     &+&  B^{Q,G}(t)\frac{i\left(P_{\mu}\sigma_{\nu \rho} + P_{\nu} \sigma_{\mu \rho}\right) \Delta^{\rho} }{4 M_N}  \nonumber \\ \nonumber
 &+& \ C^{Q.G}(t) \frac{  \Delta_{\mu} \Delta_{\nu} - g_{\mu \nu} \Delta^{2} }{M_N}
\biggr ] {u}(p)
  \ea
  one can obtain the gravitational form  factors of quarks which are related to the second  moments of GPDs
\ba
\int^{1}_{-1}dx \ x [H_q(x,\Delta^2,\xi)  = A^{q}_{2,0}(\Delta^2)+(-2\xi)^2C^{q}_{2,0}(\Delta^2), \nonumber \\
\int^{1}_{-1}dx \ x[E_q(x,\Delta^2,\xi)=B^{q}_{2,0}(\Delta^2)-(-2\xi)^2C^{q}_{2,0}(\Delta^2).
\ea
 For $\xi=0$ one has
\ba
\int^{1}_{0}dx \ x{\cal{H}}_q(x,t) = A_{q}(t); \,\ \int^{1}_{0}  dx \ x {\cal{E}}_q (x,t) = B_{q}(t).
\ea

 Our results for $A_{u+d}(t)$ are shown in Fig.13. Our GPDs with different PDFs lead
 to the same $t$ dependence of  $A_{u+d}(t)$.
 At $t=0$ these contributions equal $A(t=0) \approx 0.45$.

 The corresponding calculations for $B_q(t)$ are shown
  in Figs. 14. In this case, we have the difference at $t=0$ and some difference in the $t$ dependence
  already at small momentum transfer.
     The PDFs {\bf O12a}  give the large values (upper curve in Fig.14)
  and   PDFs {\bf GP8NNL} gave the lower values (low curve in Fig.14).  Others
        concentrated in two clusters.
  One gave  $B_{grav}(t=0) = -0.15$ (the PDFs {\bf JR8a, MRST09a, MRST09b, GJR07b}, and second gave
  $B_{grav}(t=0) = -0.11$ the PDFs {\bf ABKM09, ABM12, KKT12A, MRST02}.
  In our previous work \cite{ST-PRDGPD}, we obtained  $B_{grav}(t=0) = -0.05$ that
      is close  to the zero value.
 That is  a sort of compensation for the $u$ and $d$ quarks supporting
 the conjecture \cite{Teryaev-s3,Teryaev:2006fk} about the validity of the Equivalence Principle
 separately for quarks and gluons.

\begin{figure}
\includegraphics[width=.4\textwidth]{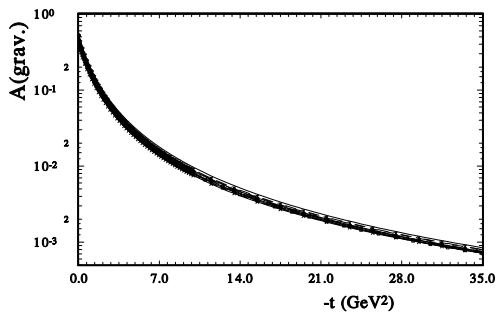} 
\caption{ Gravimagnetic form factor $A(t)$.
  }
\label{Fig_13}
\end{figure}

\begin{figure}
\includegraphics[width=.4\textwidth]{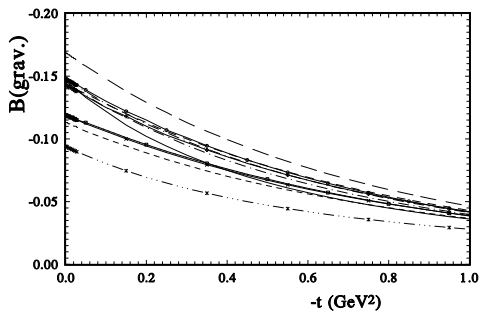} 
\caption{ Gravimagnetic form factor $B(t)$.
  }
\label{Fig_14}
\end{figure}

Note that nonperturbative analysis within the framework of the lattice OCD indicates
 that the net quark contribution to the anomalous gravimagnetic moment $B_{u+d}(0)$
 is close to zero  \cite{Gockler04,Hagler05}.
 Now, our results contradict this conclusion. Probably, it points out the important contribution of the gluon part. \\


\section{The Compton cross sections}

     The processes of the wide angle Compton scattering $\gamma^{*} p \rightarrow \gamma p$
     gave the possibility to study
     the complicated hadronic dynamics in hard exclusive processes \cite{KivVan-13}.
     There are two processes - the deeply  virtual Compton scattering (DVCS) (in this case the initial
      photon is highly virtual while the final photon is real and
      the effective masses of photons are different) and the real Compton scattering (RCS) (with both photons being real and equal). Large virtuality of the initial
      photon is sufficient for making the handbag diagram dominant \cite{R97,JiOs97}.
      The GPDs in this case have the large dependence on $\xi$.
      In the case of the RCS the GPDs  have $\xi=0$. Hence, we can use our ansatz for the  $t$ and $x$ dependence of the GPDs and calculate the corresponding cross sections.

   Our calculations are based on the
    works \cite{R98,104} and \cite{Kroll13}.
    The differential cross section for that reaction can be written as
  \ba
  \frac{d\sigma}{dt} =  \frac{\pi \alpha^{2}_{em}}{s^{2}} \frac{(s-u)^{2}}{-u s}
  &[&R_{V}^{2}(t) \ - \ \frac{t}{4 m^{2}} R^{2}_{T}(t) \\ \nonumber
    &+& \frac{t^{2}}{(s-u)^{2}} R^{2}_{A} (t)],
    \label{RCS}
\ea
  where $R_{V}((t)$, $R_{T}(t)$, $R_{A}(t)$ are the form factors given by the $1/x$
  moments of corresponding GPDs $H^{q}(x,t)$,  $E^{q}(x,t)$, $\tilde{H}^{q}(x,t)$ .
     The last is related with the axial form factors.
     As noted in \cite{Kroll13}, this factorization,
  which bears some similarity to the handbag factorization of DVCS,
  is formulated in a symmetric frame where the skewness $\xi=0$.
   For  $H^{q}(x,t)$,  $E^{q}(x,t)$ we used the PDFs obtained from the
   works \cite{JR08,KKT12,ABKM09,ABM12} with the parameters are presented in Table 5,
   obtained in our fitting procedure of the description of proton and neutron electromagnetic
   form factors. For $\tilde{H}^{q}(x,t)$ we take $\Delta q$ in the form
   \ba
   x \Delta q = N_i x^a_1 (1+ a_2 \sqrt{x} +a_3 x),
   \ea
   with the parameters are determined in \cite{80}.
     Our calculations of $R_{i}$  on the whole, correspond the calculations \cite{Kroll13},
     but the  integrals with our {\it Ansatz} of the $t$ dependence of GPDs do not divergence at momentum transfer $ -t > 2$ GeV$^2$.
     In the work \cite{Kroll13} they presented $R_{i}$ beginning from $-t=4$ GeV$^2$.
     Note that the last term in Eq.(\ref{RCS}) has the small coefficient and its impact on the differential cross sections of RCS is very small (from 2$\%$ at small $t$ and up to $10\%$
     at large momentum transfer). It is essentially less than theoretical indeterminacy.

     Our calculations of  the differential cross sections of RCS are shown  in Fig.15 at
     three energies $s = 9.8, 10.92$ and $20$. Obviously, the calculations have sufficiently
     good coincidence with the existing experimental data and in whole coincides with
     calculations \cite{Kroll13}.
      The behavior of the experimental data
     at $s=9.8$ GeV$^2$ and large $t$  is probably connected with the kinematical property when $-t
     \rightarrow s$. Probably, it is necessary to take into account the next NLO terms \cite{KivVan-13}.

\begin{figure}
\includegraphics[width=.4\textwidth]{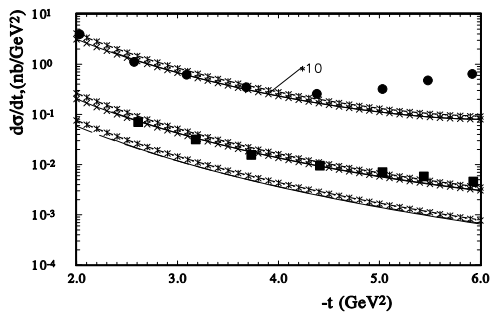} 
\caption{ Differential Compton cross sections $\gamma p \rightarrow \gamma p$;
 the curves are our calculations at $s=8.9$ GeV$^2$ ( with factor $10$), $s=10.92$ GeV$^2$,
 and $s=20$ GeV$^2$ ( hard line, dashed line, dot-dashed line with (x), med-dashed line with ($+$),
   correspond to the PDFs [50,52,53,48];
   the data points \cite{105} are for $s=8.9$ GeV$^2$ (circles with factor $10$);
  $s=10.92$ GeV$^2$ (squares).
  }
\label{Fig_14}
\end{figure}

\newpage

\section{Conclusions}

   The complex analysis of the corresponding description of the electromagnetic form factors of the proton and neutron
    by the different  PDF sets  (24 cases) was carried out . These
   PDFs include the  leading order (LO), next leading order (NLO) and next-next leading order (NNLO)
   determination of the parton distribution functions. They used the different forms of the $x$ dependence
   of  PDFs, eqs. (\ref{sq1} - \ref{ex3}). The analysis was carried out with different forms of the
   $t$ dependence of GPDs. The minimum number of  free parameters was six and maximum were ten.
   We found that the best description was given by  PDFs  \cite{ABKM09}. In this case, the increase in the number of the free
   parameters leads to a small decrease in $\chi^2$. It means that the $x$ dependence of PDFs
    corresponds sufficiently well to the $u$ and $d$ distributions in the nucleon  to reproduce the
   electromagnetic form factors. Note that these PDFs used the special power $x$ dependence of PDFs.
    The  other PDFs   \cite{JR08,ABM12,KKT12,GJR07,MRST02}
    also a similar behavior as \cite{ABKM09} and have a small change in $\chi^2$ with increasing number of the free parameters and lead to good  descriptions with minimum free parameters.
    Note, it is remarkable that old PDFs  \cite{MRST02} are  in this list too.
    Practically in all our calculations
     PDFs \cite{ABKM09} gave the medium result between other PDFs. This confirms the result  the minimum of
     $\chi^2$ obtained   with  the  minimum of  number of free parameters.
    We did not find a visible difference between PDFs with a different order.
       This is in accord  with the conclusion of  paper \cite{Forte13NLO} that the theoretical uncertainty of PDFs exceeds the uncertainty of the perturbative series.


    In the final analyses, we found that  all PDFs  in  the simultaneous description of the proton and neutron
    electromagnetic form factors led to the "polarization" case of the $t$-dependence of the
    form factors.

    The flavor dependence in these cases, in most part, comes from the spin dependence
        part of  PDFs.  We obtained good descriptions of the electric and magnetic form factors
        of the proton and neutron simultaneously. We found that  different PDFs
        gave almost the same descriptions of the proton form factors at small momentum transfer.
        The difference appears only at large $t$. Our calculations of the $u$ and $d$ quark contributions show the same $t$ dependence
        at large $t$.


      All PDFs gave approximately the same size and the $t$-dependence of the gravitation form factors $A(t)$
      as the second moment of the GPDs. The size of the gravimagnetic form factor $B(t=0)$ differs from zero.
      The PDFs \cite{ABKM09} gave $B_{grav.}(t=0) = -0.12$. It is above the result obtained by us
      in the qualitative description of the nucleon form factor \cite{ST-PRDGPD} which was  $B_{grav.}(t=0) = -0.05$.
      Hence, this may indicate on the important contribution of the gluon part.

\end{document}